\documentclass[twocolumn,secnumarabic,amssymb, nobibnotes, superscriptaddress, nofootinbib,aps,prd]{revtex4-2}
\usepackage[colorlinks,citecolor=blue,linkcolor=blue,anchorcolor=blue,filecolor=blue, urlcolor=blue]{hyperref}
\usepackage{natbib}

\usepackage{amsmath}
\usepackage{adjustbox}
\usepackage{graphicx}
\usepackage{booktabs}       % professional-quality tables
\usepackage{amsfonts}       % blackboard math symbols
\usepackage{nicefrac}       % compact symbols for 1/2, etc.
\usepackage{microtype}
\usepackage{cleveref}
\usepackage{xcolor}
\usepackage{multirow}
\usepackage{makecell}
\usepackage{array}
\usepackage{orcidlink}
\usepackage{float}

\begin{document}

\author{Adamu Issifu \orcidlink{0000-0002-2843-835X}} 
\email{ai@academico.ufpb.br}

\affiliation{Instituto Tecnológico de Aeronáutica,\\ CEP 12.228-900, São José dos Campos, SP, Brazil} 

\author{Tobias Frederico \orcidlink{0000-0002-5497-5490}} 
\email{tobias@ita.br}

\affiliation{Instituto Tecnológico de Aeronáutica,\\ CEP 12.228-900, São José dos Campos, SP, Brazil}

\title{Hot quark matter and merger remnants} 

\begin{abstract}
This work investigates hot quark matter under the thermodynamic conditions characteristic of a binary neutron star (BNS) merger remnants. We used the density-dependent quark mass model (DDQM) to access the microscopic nuclear equation of state (EoS) in a series of snapshots. The strange quark matter (SQM) is studied at finite temperature and entropy, in the presence of electrons and muons and their corresponding neutrinos to simulate the BNS merger conditions. For the first time, we introduced temperature into the DDQM model using a lattice QCD-motivated approach to construct both isentropic and isothermal EoSs. We observe that as the entropy of the SQM increases, the merger remnant becomes more massive and increases in size, whereas the neutrino abundance also increases. In the fixed-temperature case, on the other hand, we observe that the entropy spreads from the surface towards the center of the remnant. We determine the particle distribution in the core of the remnants, the structure of the remnant, the temperature profile, sound velocity, and the polytropic index, and discuss their effects. The strange-quark star (SQS) remnants satisfy the $2\,{\rm M_\odot}$ mass constraint associated with neutron stars (NS).

\end{abstract}

\maketitle

\section{Introduction}

The merger of two NSs leading to the detection of a gravitational wave in event GW170817 (GW)~\cite{LIGOScientific:2017vwq} results in the formation of massive remnants and subsequent matter ejection in the interstellar medium~\cite{Baiotti:2016qnr, Barack:2018yly}. The remnant comprises a central object with a massive disk around it. The characteristics of the disk and the central object depend on the masses of the merging stars, the magnetic field, NS spin, and their corresponding EoSs~\cite{Hotokezaka:2011dh, Bauswein:2013jpa, Hotokezaka:2013iia}. The gravitational wave was observed from the inspiral phase by Advanced LIGO~\cite{LIGOScientific:2014pky} and Advanced Virgo~\cite{VIRGO:2014yos} interferometers. This observation paved a new path into the microphysics of NSs, offering the possibility to probe matter dynamics at extreme temperature and density conditions inside these remnants. The constraint on the tidal effect of the coalescing bodies and the maximum NS mass and radii has enabled nuclear astrophysicists to impose a strict constraint on the EoS for cold NSs~\cite{LIGOScientific:2018cki}. The detection also coincided with the observation of a $\gamma$-ray burst, GRB 170817A (GRB)~\cite{LIGOScientific:2017zic, Goldstein:2017mmi}  which helps in classifying the binary contained matter using the relative time delay between the GW and GRB and the measured speed of sound~\cite{LIGOScientific:2017zic}.

Currently, gravitational waves have been observed,  and future gravitational wave observatories are expected to observe more neutron star mergers with higher confidence levels. In this light,  numerical simulations for the BNS merger are important to interpret the data and further constrain the EoS for NSs. Temperatures in the postmerger phase are expected to reach as high as $\sim 50-100$ MeV, making it suitable for finite temperature studies~\cite{Perego:2019adq}. Even though the current gravitational wave detectors have not reported on a postmerger signal yet~\cite{LIGOScientific:2017vwq, LIGOScientific:2017ync, LIGOScientific:2020aai}, it is expected that future detectors such as the Einstein Telescope~\cite{Maggiore:2019uih} and Cosmic Explorer~\cite{LIGOScientific:2016wof} detectors with improved sensitivities and high-frequency features may detect BNS postmerger signals. The current sensitivity upgrades on the existing instruments also promise a better quality BNS detection with a better sky location~\cite{KAGRA:2013rdx}.

It has been predicted through numerical relativity simulations that the final product of BNS mergers is a prompt formation of a black hole through gravitational collapse or the formation of a remnant NS~\cite{Shibata:1999wm}. In the case of NS formation, the baryon density ($n_b$) of a typical $\beta$-equilibrated matter of the individual NSs of about $n_b\sim 2-3 n_0$ ($n_0$ is the nuclear saturation density) increases to about  $n_b\sim 5-6 n_0$ with a high remnant temperature~\cite{Sekiguchi:2011mc, Bernuzzi:2015opx, Radice:2018pdn}. A typical temperature in the postmerger remnant core at the initial stages ranges from 20 to 60 MeV~\cite{Bauswein:2010dn}. This temperature further increases in the contact layer at the early stages shooting up to about 150 MeV~\cite{Bauswein:2010dn, Rosswog:2012wb}. These extreme conditions are expected to be reached on shorter time scales due to the violent nature of the event, in milliseconds, under conditions relatively different from the formation of proto-NSs (PNS) through supernova remnants~\cite{roberts2012protoneutron, Issifu:2023qoo}. Also, higher isospin asymmetries are expected in BNS merger events due to an excess of neutrons over protons in close binary systems NS-NS, and NS-BH~\cite{Shibata:2011jka, Faber:2012rw, Rosswog:2015nja}, contrary to PNSs~\cite{Prakash:1996xs, Janka:2006fh}.

Since Ivanenko and Kurdgelaideze hypothesized the existence of SQSs in 1965~\cite{Ivanenko:1965dg, Ivanenko:1969bs}, together with Bodmer \cite{PhysRevD.4.1601} and Witten \cite{Witten:1984rs} conjecture in 1971 and 1984 respectively, that SQM is likely to be the true ground state of hadronic matter, quark stars (QSs) have been intensively studied (see e.g. ~\cite{daSilva:2023okq, PhysRevD.4.1601, Bombaci:2004mt, Herzog:2011sn, Annala:2019puf}). Nowadays, there is sufficient information about compact objects through the electromagnetic spectrum and gravitational wave detection. Analysis of these observational data has led to constraints on macroscopic properties such as the mass, radii, and tidal deformability of compact astrophysical objects~\cite{LIGOScientific:2017vwq, LIGOScientific:2018cki}. Recent data obtained from the Neutron Star Interior Composition Explorer (NICER) led to simultaneous measurement of the mass and radii of PSR J0030+0451 and PSR J0740+6620 with reasonable accuracy, further constraining the composition of stellar matter~\cite{Riley:2019yda, NANOGrav:2019jur, Fonseca:2021wxt, Riley:2021pdl}. However, the exact composition of these pulsars is still a research subject. Thus, these observations in recent years have led to robust restrictions on the EoS ruling out most phenomenological quark models that are too soft to reach the 2$\rm M_\odot$ constraint \cite{Ozel:2006bv, Weber:2004kj, Albino:2021zml}. %Additionally, it has been established theoretically that, at a baryon density sufficiently high (in the inner core of a compact object), the Fermi energy of the nucleon is high enough to allow possible new degrees of freedom such as exotic baryons~\cite{Marquez:2022gmu, Issifu:2023qyi, Thapa:2021ifv, Bonanno:2011ch} and deconfined QM~\cite{Kumar:2023lhv, Issifu:2023ovi}. 
Ever since the SQM was conjectured as the true ground state of nuclear matter~\cite{Witten:1984rs, PhysRevD.4.1601, Ivanenko:1969gs, Itoh:1970uw}, it has been understood that hadronic matter can transform into QM, thereby transforming NSs into QSs. As discussed in~\cite{alcock1986strange, Gupta:2002hj, shen2005slowly, Dexheimer:2013eua, Dexheimer:2012mk, Drago:2013fsa, drago2016scenario}, proto-strange QSs with bare or dynamically irrelevant nucleonic matter crust ~\cite{kettner1995structure, usov1997low, melrose2006pair} can be formed through a BNS merger. %That notwithstanding, there is a strong argument against the presence of a crust on QSs~\cite{kettner1995structure, usov1997low, melrose2006pair}. 

In this paper, we employ the DDQM under thermodynamic conditions characteristic of the BNS merger remnant, assuming that the remnant is QS composed of self-bound QM. The DDQM is characterized by two free parameters D (a dimensionful constant representing the confining strength) and C (a dimensionless constant representing the deconfinement strength)  given through the cubic-root law \cite{Peng:1999gh, Xia:2014zaa}
\begin{equation}\label{q1}
   m_i = m_{i0} + \dfrac{D}{n_b^{1/3}}+Cn_b^{1/3},
\end{equation}
with $m_i$ the equivalent of a dressed quark mass, $m_{i0}$ the current quark mass and $n_b$ the baryon density. The last two terms in (\ref{q1}) can be associated to a Cornell-like potential for confining heavy quarks~\cite{PhysRevLett.34.369, Bali:2000gf, Afonin:2022aqt}, $V_c(r_*) = -\beta/r_*+\sigma r_*$, where $\sigma$ and $\beta$ are constants that can be determined through fitting to lattice QCD (lQCD) data. It comprises a Coulomb-like term 
dominant at short distances with strength $\beta$ from asymptotic freedom and a linear confining part with strength $\sigma$,  which is also associated with the QCD string tension. In the above expression, the average inter-particle separation distance $r_*$ can be associated with the baryon density through an ansatz $r_*\sim 1/n_b^{1/3}$. 
We choose $\sqrt{D} = 127.40\, {\rm MeV}$ and $C= 0.8$ reached in~\cite{daSilva:2023okq} through Bayesian analysis of the DDQM at $T =0\,{\rm MeV}$ to serve as a benchmark for our study. These parameters produce a stable QS according to Bodmer and Witten's conjecture which hypothesizes that the binding energy of SQM must be less than the energy per nucleon ($E/A$) of $^{56}$Fe, i.e. $(\varepsilon/n_b)_{\rm SQM}\leq 930{\rm MeV}$.  In the same vein, the two-flavor quark matter (2QM ) must simultaneously satisfy $(\varepsilon/n_b)_{\rm 2QM} > 930{\rm MeV}$~\cite{Witten:1984rs, PhysRevD.4.1601, fraga2014interacting} i.e., the $E/A$ of $^{56}$Fe should be larger for 2QM for $u$ and $d$ quarks to be more tightly bound otherwise, they would dissociate into free $u$ and $d$ quarks. 
Also, these choices lead to SQS that satisfies the maximum mass constraint of pulsars PSR J0952-0607~\cite{Romani:2022jhd}, PSR J0740+6620~\cite{Riley:2021pdl} and PSR J0030+0451~\cite{Riley:2019yda}, and others that satisfy the 2M$_\odot$ constraint. We fix the entropy per baryon, $S/n_b$ of the remnant matter, and calculate the temperature profile, particle distributions,  speed of sound, polytropic index, and the structure of the star. On the other hand, we consider a remnant matter at a uniform temperature and study the entropy distribution from the disk to the core of the remnant and the other properties of the stellar matter also investigated at fixed entropy per baryon.  

The paper is organized as follows: In Sec.~\ref{model}, we present the model that serves as the basis for this investigation. This section is divided into four subsections, in Subsec.~\ref{th1} we present the thermodynamic treatment of the model, in Subsec.~\ref{msc} we present the foundation of the mass scaling and its scaling with temperature, in Subsec.~\ref{sqm} we present the properties of the SQM under investigation, and finally, we present the thermodynamic conditions for BNS merger in Subsec.~\ref{thc}. We present our results and analyses in Sec.~\ref{ran}, and the final remarks in Sec.~\ref{fr}.

\section{The model}\label{model}

The variation of particle self-energies depending on the media in which they propagate is well-documented in the literature \cite{Huang:2020hdv, Xing:2007fb}. In the simplest case, this mass is typically referred to as the effective mass to distinguish it from the 'bare' mass. When models are expressed in terms of temperature and chemical potentials they are referred to as the quasiparticle models~\cite{Zhang:2021qhl} (see e.g.~\cite{Backes:2020fyw, Wen:2005uf, chu2021quark, Schertler:1996tq, Peshier:1999ww, Gardim:2009mt, Marzola:2022zez, Marzola:2023tdk}).

Now we focus our attention on DDQM, which is used in our study. Indeed, the DDQM is a quasiparticle model originally formulated to investigate the confining characteristics of QM~\cite{Fowler:1981rp, Plumer:1984aw} and was later on extended to calculate the EoS~\cite{Chakrabarty:1989bq, PhysRevD.51.1989, PhysRevLett.74.1276, Peng:2000ff}, SQM viscosity and r-modes dissipation~\cite{Zheng:2003hj}, diquark properties~\cite{Lugones:2002vd} and compact astrophysical objects~\cite{daSilva:2023okq, Issifu:2023qoo, Backes:2020fyw, chu2021quark}. The application of this model to study compact astrophysical objects has largely been successful in fulfilling the 2${\rm M}_\odot$ mass and radii constraints of NSs, making it a viable QS model. That notwithstanding, it is associated with two known limitations, namely: the equivalent mass scaling and thermodynamic consistency. However, these issues have been tackled by several authors~\cite{Backes:2020fyw, Chen:2021fdj, Peng:2000ff, Peng:2008ta} with some promising outcomes.

\subsection{Finite temperature formalism of DDQM}\label{th1}

The free energy density of the hot SQM system can be expressed as 
\begin{align}\label{1}
    F &= F(T,V,\{n_i\},\{m_i\})\nonumber\\
    &=\Omega_0(T, V,\{\mu^*_i\},\{m_i\}) + \sum_{i=u,d,s}\mu^*_in_i,
    \end{align}
where $\mu^*_i$ is the effective chemical potential for the $i$th particle, $m_i\rightarrow m_i(n_b, T)$ is the density and temperature-dependent particle mass, and $\Omega_0$ is the thermodynamic potential similar to the non-interacting particle system, which will be used as an intermediate quantity at this point. The thermodynamic potential $\Omega$ will be determined and presented at the end of this subsection as a function of $\Omega_0$. 

The expression for $F$ has the same form as the well-known expression for the free fermionic system with the particle mass replaced by the equivalent particle mass $m_i(n_b, T)$ and the real chemical potential $\mu_i$ replaced by $\mu^*_i$ for thermodynamic consistency. We connect $\mu_i^*$ and $\Omega_0$ to the independent state variables; $T,\; n_i$  and the volume $V$ through
    \begin{equation}\label{1a}
    n_i = -\dfrac{\partial}{\partial\mu^*_i}\Omega_0(T, V,\{\mu^*_i\},\{m_i\}),
\end{equation}
this expression is well-known in thermodynamics with $\mu_i^*$ in place of $\mu_i$ to ensure thermodynamic consistency. To determine the other thermodynamic quantities, we take a derivative of (\ref{1}), which yields
\begin{equation}\label{2}
    dF = d\Omega_0 + \sum_in_id\mu^*_i + \sum_i\mu^*_idn_i,
\end{equation}
here, $\Omega_0$ can be expressed in terms of the  dependent variables as
\begin{equation}\label{3}
    d\Omega_0 = \dfrac{\partial\Omega_0}{\partial T}dT + \sum_i\dfrac{\partial\Omega_0}{\partial\mu^*_i}d\mu^*_i + \sum_i\dfrac{\partial\Omega_0}{\partial m_i}dm_i + \dfrac{\partial\Omega_0}{\partial V}dV,
\end{equation}
and the equivalent mass as
\begin{equation}\label{4}
    dm_i = \dfrac{\partial m_i}{\partial T}dT + \sum_j\dfrac{\partial m_i}{\partial n_j}dn_j.
\end{equation}
Combining (\ref{2}), (\ref{3}) and (\ref{4}) and regrouping terms, we find that:
\begin{align}\label{4c}
    dF &= \left( \dfrac{\partial\Omega_0}{\partial T} + \sum_i\dfrac{\partial\Omega_0}{\partial m_i}\dfrac{\partial m_i}{\partial T}\right)dT \nonumber\\&+ \sum_i\left(\mu_i^* + \sum_j\dfrac{\partial\Omega_0}{\partial m_j}\dfrac{\partial m_j}{\partial n_i}\right)dn_i + \dfrac{\partial \Omega_0}{\partial V}dV.
\end{align}
Comparing the above result to the standard expression for the thermodynamic free energy density; 
\begin{equation}\label{4d}
    dF = -SdT + \sum_i\mu_idn_i + \left(-P -F + \sum_i\mu_in_i\right)\dfrac{dV}{V},
\end{equation}
where $S$ is the entropy density and $P$ is the pressure of the system, we identify:
\begin{equation}\label{4a}
    S = -\dfrac{\partial \Omega_0}{\partial T} - \sum_i \dfrac{\partial m_i}{\partial T}\dfrac{\partial\Omega_0}{\partial m_i},
\end{equation}
and
\begin{equation} \label{5}
    P = -F + \sum_i\mu_in_i - V\dfrac{\partial\Omega_0}{\partial V},
\end{equation}
where the real chemical potential is given in terms of the effective one as
\begin{equation} \label{6}
    \mu_i = \mu^*_i + \sum_j\dfrac{\partial \Omega_0}{\partial m_j}\dfrac{\partial m_j}{\partial n_i} \equiv \mu_i^* - \mu_{\rm int},
\end{equation}
where $\mu_{\rm int}$ is the chemical potential that carries the interaction. { For an infinitely large system of SQM, Eq.~(\ref{1}) is independent of the volume of the system and hence (\ref{5}) takes the form: 
\begin{align}\label{6a}
P & = -F + \sum_i \mu_in_i \nonumber \\
   & = -\Omega_0 + \sum_{i,j}n_i\dfrac{\partial \Omega_0}{\partial m_j}\dfrac{\partial m_j}{\partial n_i},
\end{align}
in the second step we have substituted (\ref{1}) in place of $F$ and (\ref{6}) in place of $\mu_i$ to obtain a result in terms of $\Omega_0$ and $\mu^*_i$. Similar expressions were derived in~\cite{Wen:2005uf, Xia:2014zaa, Chen:2021fdj} for DDQM using a thermodynamic consistent approach, and in~\cite{Peng:2000ff} using the general ensemble theory. Additionally, the last term in (\ref{5}) only persists if the finite size effect of the system cannot be overlooked whether the particle masses are fixed or not.}
The energy density of the system can then be determined from the fundamental relation with the Helmholtz free energy, namely  $\varepsilon = F - TS$, by substituting (\ref{1}) and (\ref{4a}), resulting in:
\begin{equation}\label{6b}
    \varepsilon = \Omega_0 + \sum_i\mu_i^*n_i - T\dfrac{\partial \Omega_0}{\partial T} - T\sum_i \dfrac{\partial m_i}{\partial T}\dfrac{\partial\Omega_0}{\partial m_i}.
\end{equation}
Finally, the real thermodynamic potential $\Omega$ of the system can be determined as  
\begin{equation}
    \Omega = F - \sum_i\mu_in_i = \Omega_0 - \sum_{i,j}n_i\dfrac{\partial \Omega_0}{\partial m_j}\dfrac{\partial m_j}{\partial n_i}.
\end{equation}

It should be noted that \(\Omega_0\) consists of both particle and antiparticle contributions, so the full expression is \(\Omega_0^{\pm} = \Omega_0^- + \Omega_0^+\), where \(\Omega_0^-\) and \(\Omega_0^+\) represent the particle and antiparticle contributions, respectively. Consequently, the explicit expression for the $\Omega_0^\pm$ is given as
\begin{align}\label{om}
    \Omega^\pm_0 &= - \sum_i\dfrac{g_i T}{2\pi^2}\int p^2dp\Big[\ln[1+e^{-(\epsilon_i-\mu_i^*)/T}] \nonumber\\&+ \ln[1+e^{-(\epsilon_i+\mu_i^*)/T}]\Big],
\end{align}
where $i = u,\;d,\;s,\;e,\;\mu,\; \nu_\mu, \;\text{and}\; \nu_e$ (up, down, strange, electron, muon, muon neutrino, and election neutrino respectively) represents the quarks and leptons present in the system and $g_i$ the degeneracy factor. The number  density of each fermion species is also given by two terms, $n_i^\pm = n^+_i -  n_i^-$ thus,
\begin{align}
    &n_i^{\pm}=-\dfrac{\partial \Omega_0^\pm}{\partial\mu^*_i} \nonumber\\&= \sum_i\dfrac{g_i}{2\pi^2}\int_0^\infty  p^2dp\Big[\dfrac{1}{1+e^{(\epsilon_i-\mu_i^*)/T}} - \dfrac{1}{1+e^{(\epsilon_i+\mu_i^*)/T}}\Big].
\end{align}
The derivatives in $S,\;\mu_i^*,\;P, \text{and},\;\varepsilon$ can be written explicitly as
\begin{align}
    \dfrac{\partial\Omega_0^\pm}{\partial m_i} =& \sum_i\dfrac{g_im_i}{2\pi^2}\int_0^\infty \Big[\dfrac{1}{e^{(\epsilon_i-\mu^*_i)/T}+1}\nonumber\\&+\dfrac{1}{e^{(\epsilon_i+\mu^*_i)/T}+1}\Big]\dfrac{p^2dp}{\epsilon_i},
\end{align}
and
\begin{align}
  \dfrac{\partial\Omega_0^\pm}{\partial T} &= -\dfrac{g_i}{2\pi^2}\int_0^\infty\Big(\ln\left[1+e^{-(\epsilon_i - \mu^*_i)/T}\right] \nonumber\\&+ \dfrac{(\epsilon_i - \mu_i^*)/T}{1+e^{-(\epsilon_i - \mu^*_i)/T}} + \ln\left[1+e^{-(\epsilon_i + \mu^*_i)/T}\right] \nonumber\\&+ \dfrac{(\epsilon_i + \mu_i^*)/T}{1+e^{-(\epsilon_i + \mu^*_i)/T}}\Big)p^2dp,
\end{align}
with $\epsilon_i = \sqrt{p^2+m_i^2}$. %Similar derivations can be found in Refs.\cite{Wen:2005uf, Peng:2000ff, Xia:2014zaa}.

\subsection{Mass scaling with temperature}\label{msc}
We will adopt the cubic-root equivalent mass scaling formula, which was originally derived for \( T = 0 \) \cite{Peng:1999gh, Xia:2014zaa} and later extended to finite temperature scenarios in \cite{Wen:2005uf, Chen:2021fdj}. The derivation of quark mass scaling and its thermodynamic consistency have been elaborated in \cite{Wen:2005uf, Peng:1999gh, daSilva:2023okq}. We do not intend to repeat the derivation here, but we will highlight the parts fundamental to this work, and focus more on how we intend to introduce temperature into the system. We start with the Hamiltonian for the effective quark degrees of freedom
\begin{equation}
    H_{\rm Q} = H_k + \sum_{q = u,\,d,\,s}m_{i0}\bar{q}q + H_I,
\end{equation}
with $H_k$ the kinetic term and $H_I$ the interaction term. %Terms that break the flavor symmetry have not been included. 
We now introduce an equivalent Hamiltonian density with the interacting term absorbed into an equivalent mass $m_i$ in the form 
\begin{equation}
    H_{\rm eqv} = H_k + \sum_{q = u,\,d,\,s}m_{i}\bar{q}q,
\end{equation}
hence, 
\begin{equation}
    m_i = m_{0i} + m_I,
\end{equation}
thus, to keep the energy of the system of free quarks constant, its mass must be $m_i$. To determine $m_i$, we impose the condition that $H_{\rm eqv}$ and $H_{\rm Q}$ must have the same eigenvalues for the same eigenstates, $|n_b,T\rangle$ i.e.,
\begin{equation}
    \langle n_b,T|H_{\rm Q}|n_b,T \rangle = \langle n_b,T|H_{\rm eqv}|n_b,T \rangle,
\end{equation}
applying the same equality to the vacuum state $|0\rangle$ and taking the difference we obtain
\begin{equation}\label{vf}
    m_i = m_{i0} + \dfrac{\langle H_I \rangle_{n_b,T} - \langle H_I \rangle_0}{\sum_{q}\Big[\langle \bar{q}q \rangle_{n_b,T} - \langle \bar{q}q \rangle_0\Big]},
\end{equation}
therefore $m_I$ can be identified with,
\begin{equation}\label{v1}
    m_I = \dfrac{\langle H_I \rangle_{n_b,T} - \langle H_I \rangle_0}{\sum_{q}\Big[\langle \bar{q}q \rangle_{n_b,T} - \langle \bar{q}q \rangle_0\Big]}.
\end{equation}
We remark that the interacting part of the equivalent quark mass, $m_I$, depends on the interaction energy density, and the relative quark condensates (the denominator of the above expression), which are assumed to be flavor independent.
%For color confinement and asymptotic freedom, $m_I$ must satisfy 
% \begin{equation}\label{va}
% \lim_{n_b \rightarrow 0} m_I \rightarrow \infty;\qquad{\text{color confinement condition}}\qquad
% \end{equation}
% and 
% \begin{equation}\label{v}
% \lim_{n_b \rightarrow \infty} m_I \rightarrow 0;\qquad{\text{asymptotic freedom condition.}}\qquad
% \end{equation}
The proposed structure of the equivalent mass as derived in \cite{Peng:1999gh} in terms of $n_b$ is
\begin{equation}\label{d}
    m_i = m_{i0} + \dfrac{D}{n_b^z},
\end{equation}
where $z>0$ and $D$ is a constant associated with the {vacuum energy density, $m_i(n_b)$ and $m_I(n_b)$ are functions of the baryon density (it will be shown explicitly below). This expression incorporates the confinement mechanism in the way proposed by Pati and Salam \cite{Pati1977}. They viewed confinement as a scenario in which a quark has a small mass inside a hadron and an infinitely large mass outside, making it impossible for the vacuum to support the quark outside the hadron. This can be modeled by assuming that the mass of an isolated quark is infinitely large. Similar boundary conditions are employed in the MIT bag model, where quarks have a vanishing mass inside the bag and an infinitely large mass outside the bag \cite{thomas1984chiral}. A similar view is used in the modeling of quark matter phenomenologies such as the quark mass density-dependent models \cite{Chakrabarty:1989bq, PhysRevD.43.627, 1984PhLB..139..198P, Lugones:2022upj} to study confinement \cite{Chakrabarty:1989bq} and the DDQM under consideration. The DDQM offers two advantages: it recovers the asymptotically free behavior predicted by QCD at sufficiently high \( n_b \rightarrow \infty \), and it also naturally restores the expected dynamical quark confinement at low \( n_b \rightarrow 0 \).} 

Assuming the quarks are in a confined state, the energy density of the interaction depends on both the particle density and the linear confining potential, \( V_c(r_*, T) \), which is a function of the average distance between the particles, \( r_* \), and the temperature, \( T \). Explicitly,
\begin{equation}\label{v2}
    \langle H_I \rangle_{n_b,T} - \langle H_I \rangle_0 = 3n_bV_c(r_*, T),
\end{equation}
{where factor 3 is due to the number of quark flavors participating in the strong interaction.
%, $r_*$ is the average inter-particle distance within which confinement can be observed, and $V_c$ is the linear confining potential. 
The  average quark-quark distance is related to $n_b$ by the ansatz}
\begin{equation}\label{v4}
    r_* \sim \dfrac{1}{n_b^{1/3}},
\end{equation}
we have adopted the cubic-root law where $z$ was determined to be $1/3$ in~\cite{Peng:1999gh, Xia:2014zaa, Wen:2005uf}. The confining potential in lQCD is well-known to be proportional to \( r_* \) in the form \( V_c(r_*, T) = \sigma(T) r_* \), where \( \sigma \) is the proportionality constant, commonly referred to as the string tension \cite{Glozman:2007tv, Issifu:2022pif, Pasechnik:2021ncb}.
% with this expression, (\ref{v}) and (\ref{va}) are satisfied;
% \begin{equation}
%     \lim_{n_b \rightarrow \infty}V_c(r_*,T) \rightarrow 0,
% \end{equation}
% {which expresses the asymptotic freedom and, on the other hand, the low-density limit:
% \begin{equation}
%     \lim_{n_b \rightarrow 0}V_c(r_*,T) \rightarrow \infty,
% \end{equation}
% represents color confinement.} 

The potential $V_c$ has been widely investigated as a function of temperature in the literature -- see, e.g., Refs.~\cite{Bicudo:2010hg, Kaczmarek:1999mm, Digal:2003jc} --, it was found in all cases that the string tension decreases with temperature and vanishes at $T=T_C$ ($T_C$ is the critical temperature), i.e., $\sigma(T=T_C) = 0$. Hence, the normalized $\sigma(T)$ can generally be expressed as 
\begin{equation}
    \dfrac{\sigma(T)}{\sigma_0} = \left[1-\dfrac{T^2}{T_C^2}\right]^\alpha, 
\end{equation}
where $\alpha$ is a constant usually determined in the model framework and $\sigma_0$ is the string tension determined at $T=0$. We adopt the case in which $\alpha = 1$ as applied in~\cite{PhysRevC.65.035202} to the density and temperature-dependent quark mass in the framework of the MIT bag model. In addition, a phenomenological study of quark and glueball confinements in~\cite{Issifu:2020gxy} arrived at a similar outcome. {Considering that the denominator of (\ref{v1}) must satisfy
\begin{equation}
    \lim\limits_{n_b \rightarrow 0}\dfrac{\langle\bar{q}q\rangle_{n_b}}{\langle\bar{q}q\rangle_0} = 1,
\end{equation}
in the vacuum, we can use the Taylor series to expand it to vanishing $n_b$, thus;
\begin{equation}
    \dfrac{\langle\bar{q}q\rangle_{n_b}}{\langle\bar{q}q\rangle_0} = 1-\dfrac{n_b}{n_q^*}+ {\rm higher\; order\; in \;n_b}+\cdots,
\end{equation}
where $n_q^*$ is the constant quark condensate in the vacuum. Therefore, 
\begin{equation}\label{dd}
    \sum_{q}\Big[\langle \bar{q}q \rangle_{n_b} - \langle \bar{q}q \rangle_0\Big] = \sum_{q}\Big[ - \langle \bar{q}q \rangle_0/n_q^*\Big]n_b =An_b,
\end{equation}
where $A$ is a constant incorporating the constant quark condensate.}
% is compared to the chiral condensate in the QCD vacuum in a simplified form. Even though the general expression must include density and temperature dependence, for simplicity without loss of generality, we will consider only density-dependent and use the well-known model-independent quark condensate in nuclear matter~\cite{PhysRevLett.67.961, PhysRevC.45.1881},
% \begin{equation}\label{v3}
%     \dfrac{\langle\bar{q}q\rangle_{n_b}}{\langle\bar{q}q\rangle_0} = 1-\dfrac{n_b}{n_b^*},
% \end{equation}
% where $n_b^*$ is related to the pion mass $M_\pi$, the pion decay constant $f_\pi$, and the pion-nucleon sigma term $\sigma_N$ in a form,
% \begin{equation}
%     n^*_b=\dfrac{M_\pi^2f_\pi^2}{\sigma_N}.
% \end{equation}
% Substituting (\ref{v2}) and (\ref{v3}) into (\ref{v1}) and rescaling D as
% \begin{equation}
%     D\sim - \dfrac{ 3\sigma_0 n_b^*}{ \sum_q\langle\bar{q}q\rangle_0}, 
% \end{equation}
Therefore, 
\begin{equation}
    m_I = \dfrac{D}{n_b^{1/3}}\left[1 -\dfrac{T^2}{T^2_C} \right].
\end{equation}
where 
\begin{equation}\label{dd1}
    D = \dfrac{ 3\sigma_0 }{ A},
\end{equation}
the full expression in (\ref{vf}) becomes
\begin{equation}\label{m2}
    m_i = m_{i0}+\dfrac{D}{n_b^{1/3}} \left[1 -\dfrac{T^2}{T^2_C} \right].
\end{equation}
In this expression, the interacting part vanishes when $T=T_C$, similar to the QCD string-breaking scenario. {Although we use the Hamiltonian representation to derive this expression, a Lagrangian formalism was employed to obtain a similar expression at \( T = 0 \) in \cite{Kaltenborn:2017hus} for the study of hybrid neutron stars. The extension of this formalism to finite temperature using the path integral approach has also been discussed in \cite{Ivanytskyi:2022oxv}. This expression contributes to a negative pressure through $P_i \sim n_b^2\partial m_i(n_b)/\partial n_b$, which softens the EoS, leading to a reduction in the maximum stellar mass, preventing the star from reaching the maximum required mass before its eventual collapse. This shortfall was later addressed in \cite{Xia:2014zaa}, where the author introduced a third term proportional to \( n_b \) (the term \( Cn_b^{1/3} \) in (\ref{q1})), which contributes a positive pressure to counterbalance the negative pressure, ensuring that the star remains stable as its mass increases.}

Naively, comparing (\ref{q1}) to the well-known Cornell potential \( V(r_*) = -\beta/r_* + \sigma r_* + V_0 \), where \(\beta\) is the single-gluon exchange strength and \(V_0\) is the quark self-energy, captures both the confinement and asymptotically free behavior \cite{Eichten:1978tg}. The third term in (\ref{q1}) can be correlated with the single-gluon exchange term and $V_0 \sim m_{i0}$. In this case, the sign of \(C\) becomes crucial, as a positive \(C\) indicates repulsive interactions between the quarks, while a negative \(C\) implies attractive interactions. These differences in interaction lead to distinct physical consequences for the system. A negative \(C\) was found in \cite{Chen:2021fdj}, which focuses on strangelets with lower gravitational masses. However, to study stars with higher gravitational masses within the 2 \(\rm M_\odot\) threshold, it was established that a positive \(C\) is preferable \cite{daSilva:2023okq, Backes:2020fyw}. A positive \(C\) is similar to the interactions observed in multi-quark systems \cite{Huang:2023jec, Martens:2006ac} or in quark-antiquark pairs within certain color configurations \cite{Wang:2023jqs}. This attractive interaction has been determined to be suitable for obtaining higher maximum stellar masses in other phenomenological quark models \cite{rodrigues2010quark, Albino:2021zml,alford2007quark,franzon2012self,klahn2015vector, Lopes_2021,song2019effective,otto2020nonperturbative}.

The Cornell potential has been widely explored in QCD-inspired phenomenology and lQCD and has been proven to show characteristics consistent with the behavior of QCD matter.
The full expression for the finite DDQM becomes
\begin{equation}\label{m3}
    m_i = m_{i0}+\dfrac{D}{n_b^{1/3}} \left[1 -\dfrac{T^2}{T^2_C} \right] + Cn_b^{1/3} \left[1 -\dfrac{T^2}{T^2_C} \right]^{-1}.
\end{equation}
From (\ref{q1}), the confining strength is represented by \(D\), while the Coulomb-like strength is represented by \(C\). Their values can be determined through a stability analysis of the resulting QS at \(T=0\). In quenched  lQCD, the \(T_C\) has been determined to be approximately \(T_C \sim 270 \, \text{MeV}\) \cite{Fukushima:2003fw, Laermann:2003cv, Guenther:2020jwe, Aoki:2006br}. 

To obtain the stellar properties relevant to this study, we choose a positive value for $C$, as discussed in Eq.~(\ref{q1}). The positive \( C \) in our model phenomenologically accounts for repulsive effects arising from non-singlet color channels, such as the color-octet and diquark anti-triplet, which become relevant at high densities, as discussed in~\cite{Issifu_2025, Fukushima:2010bq}. While one-gluon exchange is attractive in the color-singlet channel (favoring \( C < 0 \)), in the deconfined quark-gluon plasma phase, this interaction becomes screened \cite{Casalderrey-Solana:2011dxg} and subdominant. Studies in perturbative QCD and color superconductivity \cite{Baym:2017whm} suggest that repulsive non-singlet interactions dominate at high density. In this context, the effective sign of the interaction term in the mass formula naturally flips to \( C > 0 \). This interpretation is also supported by the Bayesian analysis in~\cite{daSilva:2023okq}, where the model parameters were constrained using astrophysical data requiring a maximum NS mass of \( 2\,M_\odot \). This constraint consistently pushed the fit toward positive values of \( C \), to generate sufficient pressure for stability.

Finally, we emphasize that our effective mass ansatz does not treat color channels individually, but encodes their net effect in a thermodynamically consistent way. The singlet channel is not neglected, but becomes subdominant due to color screening and deconfinement. Thus, the choice \( C > 0 \) captures the effective repulsion required to reconcile our EoS with observations of large NS masses and tidal deformabilities~\cite{daSilva:2023okq}.

\subsection{Properties of the SQM}\label{sqm}
The SQM is, as usual, composed of up ($u$), down ($d$), and strange ($s$) quarks with leptons (electrons $e$ and muons $\mu$ with their corresponding neutrinos $\nu_e$ and $\nu_\mu$ respectively) in $\beta$- equilibrium. Due to the weak interactions, such as $d, s\leftrightarrow u + e + \bar{\nu}_e$, $s+u \leftrightarrow u + d$, etc., the $\beta$-equilibrated matter are required to satisfy the chemical equilibrium condition,
\begin{equation}
    \mu_d^*=\mu_s^*=\mu_u^*+\mu_l-\mu_{\nu_l},
\end{equation}
where the subscripts represent the particles present, $\mu_l = \{\mu_e\;, \mu_\mu\}$ and $\mu_{\nu_l}=\{\mu_{\nu_e},\;\mu_{\nu_\mu} \}$. It is important to mention here that the real chemical potentials $\mu_i$ satisfy the same chemical equilibrium relation as the effective ones above (see Ref.~\cite{Backes:2020fyw}). Charge neutrality of the system requires that 
\begin{equation}
    \dfrac{2}{3}n_u - \dfrac{1}{3}n_d-\dfrac{1}{3}n_s-n_e-n_\mu =0,
\end{equation}
here, $n_i$ is the particle number density and the subscript $i$ represents the identity of the particle present. The baryon number is also conserved through 
\begin{equation}
    n_b = \dfrac{1}{3}\big( n_u + n_d + n_s \big) = \dfrac{1}{3}\sum_in_i.
\end{equation}
At finite temperature where neutrinos are trapped in the SQM, we fix separately the lepton numbers; $Y_{L,e} = Y_e + Y_{\nu_e}$ and $Y_{L,\mu} = Y_\mu + Y_{\nu_\mu}$, with $Y_{L,e}=(n_e+n_{\nu_e})/n_b$ and $Y_{L,\mu}=(n_\mu + n_{\nu_\mu})/n_b$ for electron and muon families, respectively. The thermodynamic conditions are based on the astrophysical phenomena under consideration. It is known that neutrinos are trapped in stellar matter when their mean free path is less than the size of the system (typically the radius of the stellar object) \cite{Alford:2019qtm, Alford:2019kdw}. Also, the electron and muon neutrino spheres are not identical hence, the trapping regime might be related to a family of leptons. Therefore, we fix the lepton numbers separately for each family of leptons and ignore the $\tau$-leptons, assuming they are too heavy to be relevant.

The quark flavors have degeneracy of $g_i = 6$ (3 colors $\times$ 2 spins), the $e$, the $\mu$ have a degeneracy of $g_i=2$ and the neutrinos have degeneracy of $g_i = 1$. The current quark masses for $u,\; d$, and $s$ used are 2.16 MeV, 4.67 MeV, and 93.4 MeV, respectively, as reported in the PDG \cite{ParticleDataGroup:2022pth}. We choose a critical temperature, $T_C = 270$MeV, the strength of the linear confinement, $\sqrt{D}= 127.40$MeV, and the dimensionless single-gluon-exchange strength, $C = 0.8$. The C and D values were taken from \cite{daSilva:2023okq} determined at $T=0$ using a Bayesian study and stability analysis to serve as a reference for our analysis. The EoS and the other remnant properties were determined by either fixing $T$ and calculating $S/n_b$ or fixing $S/n_b$ and calculating the temperature profile in addition to the other properties.

\subsection{Thermodynamic conditions for the BNS merger remnants}\label{thc}
The thermodynamic conditions such as the lepton fraction, entropy per baryon, and temperature were derived from BNS simulation results \cite{Perego:2019adq, Endrizzi:2019trv, Baiotti:2016qnr, Fields:2023bhs, Oertel:2016bki, Fields:2023bhs}. The degrees of freedom considered in BNS merger simulations are nucleons, nuclei, electrons, and positrons. In addition to that, some include hyperons, quarks, $\Delta$-isobars, kaons, other particle degrees of freedom of particles and/or neutrinos \cite{Sekiguchi:2011zd, Bernuzzi:2015opx, Foucart:2015gaa, Most:2018eaw, Radice:2021jtw}. However, most state-of-the-art simulations of binary NS mergers do not include muons, even though the exotic temperature and density conditions in the BNS merger event favor their generation. A recent simulation of BNS postmerger remnants \cite{Loffredo:2022prq}, considering muons, shows that their presence significantly affects the EoS. In this work, we include muons and their neutrinos in our numerical codes to study their effect on the EoS, particularly, we pay attention to the production of $\nu_e$ and $\nu_\mu$ as the merger remnant evolves, assuming it does not form a black hole promptly.

Muons are known to play a crucial role in the microphysics of cold NSs and are expected to be significant in neutron star mergers, where the thermodynamic conditions are more favorable for their production. {The two NSs that participated in the merger event were originally in a quasi-circular orbit until the emission of GWs caused their merger -- see a review on this subject in \cite{Baiotti:2016qnr}}. In the merging process, the two stars are compressed a few times $n_0$ ($n_0=0.152\,$fm$^{-3}$ is the nuclear saturation density \cite{Horowitz:2020evx}) and heated substantially, resulting in supersonic velocities and shock wave formation. This leads to a sharp rise in temperature and localized entropy production. Given that the ejecta from a BNS merger originates from the crust and outer core of the two neutron stars, the initial lepton fraction is expected to be low. The stellar matter of the two cold neutron stars has lepton fractions in the range of \( 0.00 \leq Y_{L,l} \leq 0.2 \) \cite{Rosswog:2012wb, Sekiguchi:2015dma}. We adopted \( Y_{L,e} = Y_{L,\mu} = 0.1 \) as the working assumption for our analysis, similar to the approach taken in the study of a merger event in hadronic stars in \cite{Sedrakian:2021qjw, Alford:2019kdw}.

The cores of the NSs are not significantly shocked at merger, so the entropy per baryon of the remnant core is estimated to be $S/n_b \lesssim 2k_B$ ($k_B$ is the Boltzmann constant). Here, we chose $S/n_b = 1\,k_B$ for our analysis. When prompt black hole (BH) formation occurs, the apparent horizon eliminates the high-density region, leaving behind a relatively cold disk with a slightly higher lepton fraction, \( Y_{L,e} = Y_{L,\mu} = 0.25 \). At this stage, the entropy of the remnant also increases, marking the termination of the evolution process. Consequently, we selected \( S/n_b = 1.5 \, k_B \) for our analysis.  As the action of the inspiral arms on the innermost part of the disk increases the entropy of the remnant, the temperature also rises accordingly. The estimated entropy at this stage is approximately \( 2 < S/n_b [k_B] \lesssim 10 \). However, for our analysis, we chose a lower limit of \( S/n_b = 3 \, k_B \) and an upper limit of \( S/n_b = 6 \, k_B \). In contrast to proto-neutron \cite{Pons:1998mm} or proto-strange \cite{daSilva:2023okq} stars, which typically exhibit moderate entropy levels, merger remnants can develop considerably higher $S/n_b$ due to shock heating and neutrino trapping in the aftermath of the merger \cite{Hanauske:2016gia}. In \cite{Most:2022wgo}, the authors examined the thermodynamic conditions of BNS mergers and heavy ion collisions to explore their similarities and differences. In this study, the flow of entropy from these two events was a crucial factor.

Additionally, we consider QM at a uniform temperature, setting a lower limit of \( T = 5 \, \text{MeV} \) and an upper limit of \( T = 50 \, \text{MeV} \). Although there is a wide spread of $T$ and $S/n_b$ at the initial stages of the remnant's evolution, these thermodynamic conditions become homogeneous in the remnant matter at the latter stages of its evolution. It has been established that the nonhomogeneous nuclear matter phase disappears at \( T_C \sim 15 \, \text{MeV} \), marking the liquid-gas phase transition. It is important to note that this critical temperature is model-dependent; however, an average value of \( T_C = 15 \, \text{MeV} \) was obtained in \cite{Shen:1998gq, Haensel:2007yy}. We choose $T=5\, {\rm MeV}$ as the lower limit because it is the threshold at which stellar matter is expected to start trapping neutrinos.

Moreover, matter under extreme conditions of temperature and densities above \( n_0 \) is also formed during core-collapse supernova (CCSN) explosions, in addition to the remnants of BNS mergers. This process leads to the formation of PNSs, which undergo several evolutionary stages before ultimately becoming either NS or a BH. The difference between the remnants of CCSN and  BNS mergers lies in the isospin asymmetry; the latter involves a higher isospin asymmetry due to an excess of neutrons over protons in NS-NS or NS-BH merger events (see \cite{Oertel:2016bki} for a review). In the case of CCSN, the core of the massive star traps neutrinos, particularly \( \nu_e \), which heats and expands the core before the explosion occurs. This process increases the proton fraction during the initial stages of the star's birth. However, after the core bounce, deleptonization occurs, leading to a reduction in the proton fraction in the PNS as it cools down to form a cold NS \cite{Prakash:1996xs, Prakash:2000jr}. In the case of a BNS merger, the proton fraction in the two cold NSs that initially take part in the merger is small. 

In recent years, there has been significant effort in theoretical work, laboratory experiments, and astrophysical observations to constrain thermodynamic properties and chemical composition of stellar matter under the conditions present in CCSN and BNS mergers. This is essential due to the connection between the macroscopic structure of the remnants during their evolution and the fundamental interactions among their constituents at the microscopic level. These factors motivate the study of such objects, as they contribute to the development of general-purpose EoSs and challenge our understanding of nature across all scales.

\subsection{Numerical Approach}
{Thermodynamic implications such as microphysical transport effect and viscosity effects caused by Urca processes are possible to cause out-of-equilibrium dynamics during BNS merger \cite{Most:2022yhe, Alford:2017rxf}. Even though these properties can contribute to small changes in gravitational wave emissions, their impact on the thermodynamic properties is quite negligible.} That notwithstanding, they can cause a fundamental adjustment in the isospin asymmetry and by extension the particle fraction in the dense matter \cite{Most:2022yhe}, which we overlook in our calculations. More importantly, thermodynamics is perfectly able to capture the entropy production in the compression phase by way of  Rankine-Hugonist shock junction condition \cite{rezzolla2013relativistic}. Additionally, it has been shown in \cite{Most:2022yhe} that the viscous entropy production in the BNS merger is small in the order of $0.1/$baryon [in natural units] which we consider negligible in our calculations. %Here, we 

Determining the structure of the BNS merger remnant is quite complicated due to the presence of nonzero net angular momentum, essentially concentrated in the disk, and is gravitationally bound. Additionally, the thermodynamic properties of the core in the local rest frame are crucial for providing valuable insights into the characteristics of the remnant throughout its evolution \cite{Most:2022wgo}. Furthermore, since the physical viscosity of the remnant core is neglected and the flow is considered isentropic, the entropy per baryon, along with \(\beta\)-equilibrium \cite{Ardevol-Pulpillo:2018btx, Hammond:2022uua} and a fixed lepton fraction, can serve as effective tools for describing the evolution of the remnant through a series of snapshots \cite{Sedrakian:2022kgj}.

\section{Results and analysis}\label{ran}
\begin{table*}[ht]
\begin{center}
\begin{tabular}{ |c|c| c| c| c| c| c| c|}
\hline 
  \multicolumn{7}{|c|}{fixed entropy} \\
  
\hline
  S$/$n$_b[k_B]$&Y$_{L,l}$ & M$_{\rm max}[{\rm M}_{\odot}]$ & $R[{\rm km}]$ &$\varepsilon_0[{ MeV fm^{-3}}]$& $n_c$ [$fm^{-3}$]& $T_c$ [MeV]\\
 \hline 

 % \multicolumn{6}{c|}{Country List} \\
 % \hline
 1& $Y_{L,e}=Y_{L,\mu}=0.1$ & 2.254 & 14.38 & 658 & 0.57 & 9.02\\
1.5& $Y_{L,e}=Y_{L,\mu}=0.25$ & 2.35 & 14.83 & 616.62 & 0.51 & 12.21\\  
3& $Y_{L,e}=Y_{L,\mu}= 0.1$ & 2.29 & 14.46 & 653.40 & 0.55 & 27.36\\
6& $Y_{L,e}=Y_{L,\mu}= 0.1$ & 2.53 & 15.74 & 538.67 &0.43 &53.11\\
\hline 
  \multicolumn{7}{|c|}{Fixed Temperature} \\
  \hline
  T[MeV]&Y$_{L,l}$ & M$_{\rm max}[{\rm M}_{\odot}]$ & $R[{\rm km}]$ &$\varepsilon_0[{ MeV fm^{-3}}]$& $n_c$ [$fm^{-3}$]& $T_c$ [MeV]\\
 \hline
5& $Y_{L,e}=Y_{L,\mu}=0.1$ & 2.25 & 14.31 & 679.76 & 0.59 & ---\\
50& $Y_{L,e}=Y_{L,\mu}=0.1$ & 2.55 & 16.65 &514.55& 0.41 & ---\\
\hline \hline
 
\end{tabular}
\caption{BNS remnant properties. Here, $Y_{L,l}$ is the lepton fraction, M$_{max}$ is the maximum mass, $R$ radius, $\varepsilon_0$ the central energy density, $n_c$ is the central baryon density and $T_c$ is the core temperature.} 
\label{T1}
\end{center}
\end{table*}

% \begin{center}
% \begin{tabular}{ |c|c|c| } 
% \hline
% col1 & col2 & col3 \\
% \hline
% \multirow{3}{4em}{Multiple row} & cell2 & cell3 \\ 
% & cell5 & cell6 \\ 
% & cell8 & cell9 \\ 
% \hline
% \end{tabular}
% \end{center}
In Table~\ref{T1}, we show the lepton fraction $Y_{L,l}$, the maximum mass ${\rm M_{max}}$, its radii $R$, entropy per baryon $S/n_b$, core temperature $T_c$, {and the central energy and baryon densities, $\varepsilon_0$ and $n_c$, respectively. We observe that the ${\rm M_{max}}$ and $R $ increase by increasing $S/n_b$, while $\varepsilon_0$ and $n_c$ decrease} causing a rise in the core temperature of the NS remnant matter. In the case of an increase in $Y_{L,l}$ and $S/n_b$ (condition for prompt formation of a black hole) the ${\rm M_{max}}$, $R$ and $T_c$ increases with a decrease in $\varepsilon_0$ and $n_c$. For uniform-temperature matter, a hotter QM has higher ${\rm M_{max}}$ and $R$ while $\varepsilon_0$ and $n_c$ decrease, as usual. {Naturally,  the ratio $\varepsilon_0/n_c$  increases as $S/n_B$ rises and $T_c$ increases.} It has been established (see \cite{Hotokezaka:2013iia, Perego:2019adq}) that a typical density in BNS merger remnant is between $2n_0$ and $6n_0$, this agrees with the values of $n_c$ on Table~\ref{T1} which ranges between $2.7n_0$ and $4n_0$. Following the tabulated results for fixed $S/n_b$ and $T_c$, hotter remnants accrete more mass and increase in size.

\begin{figure*}[ht!]
  %\centering
  \includegraphics[scale=0.5]{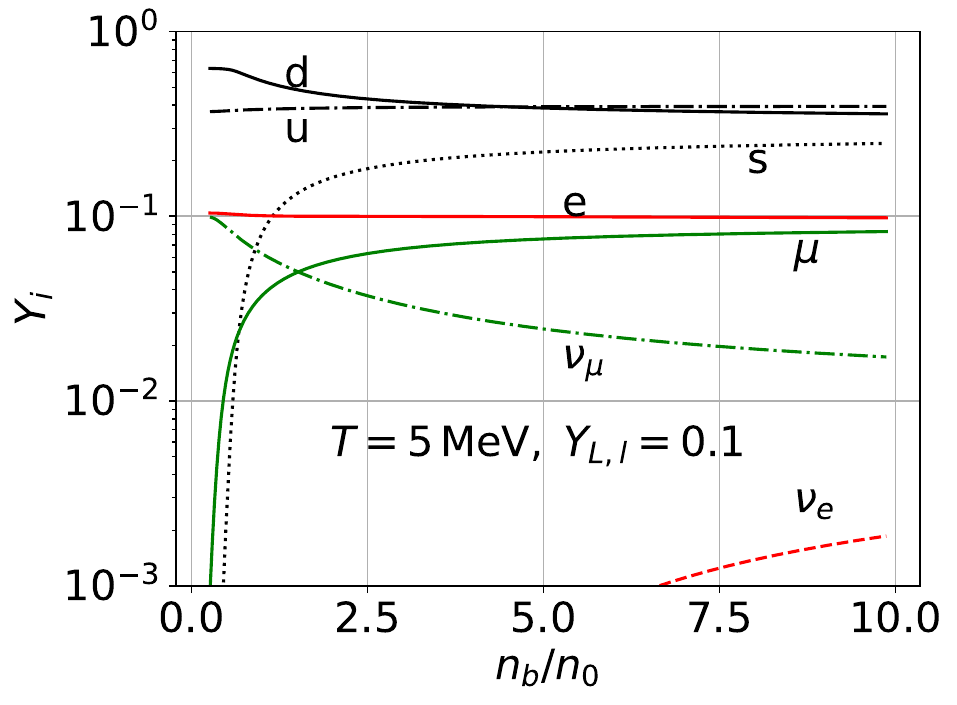}
  \quad
   \includegraphics[scale=0.5]{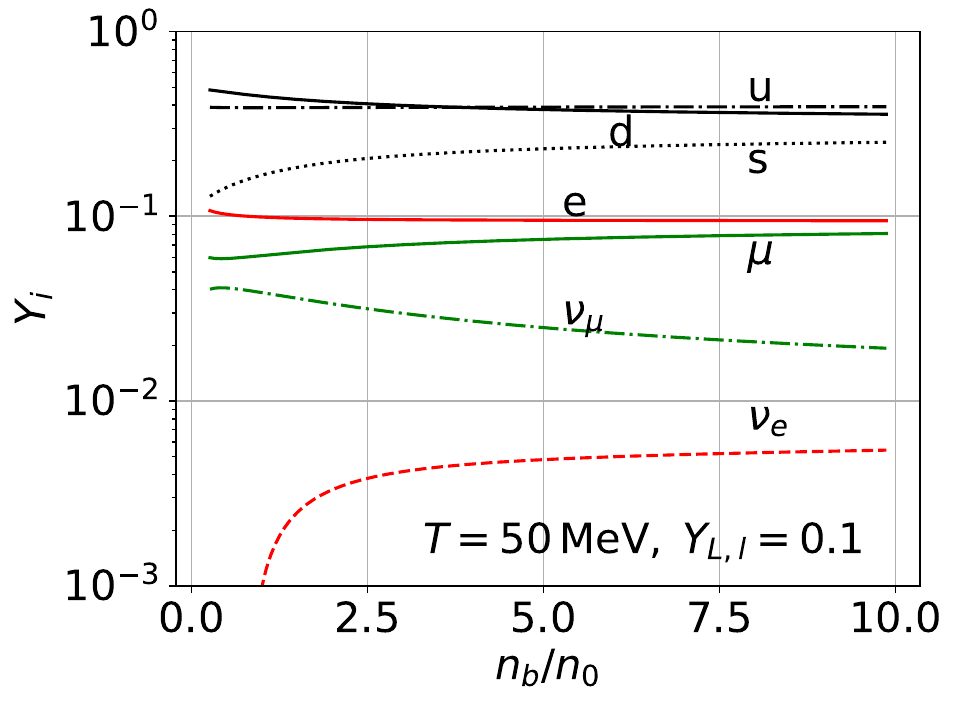}
   \quad
   \includegraphics[scale=0.5]{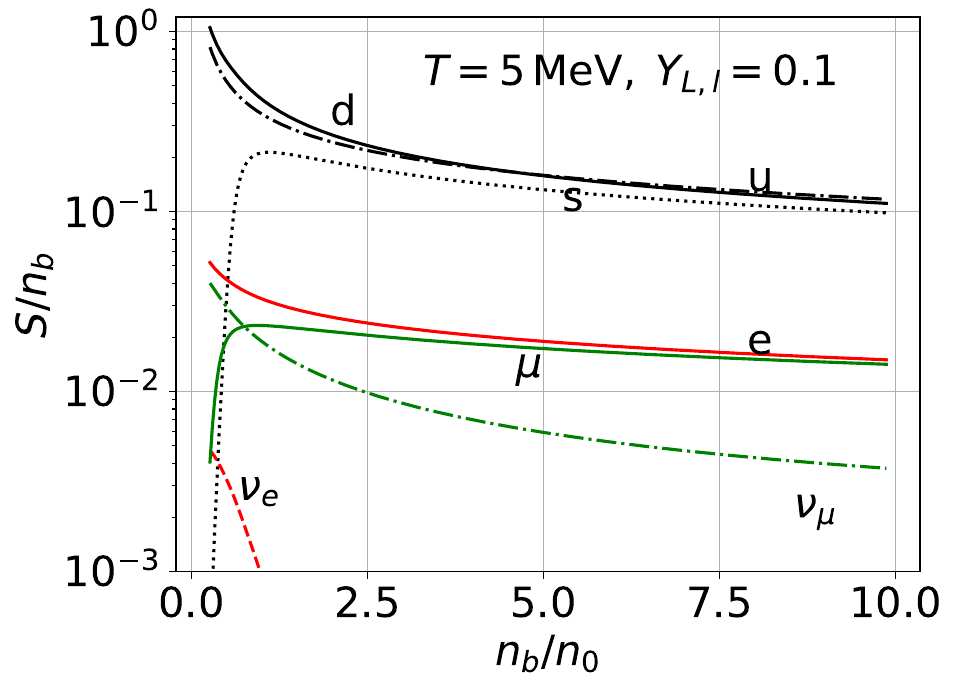}
   \quad
  \includegraphics[scale=0.5]{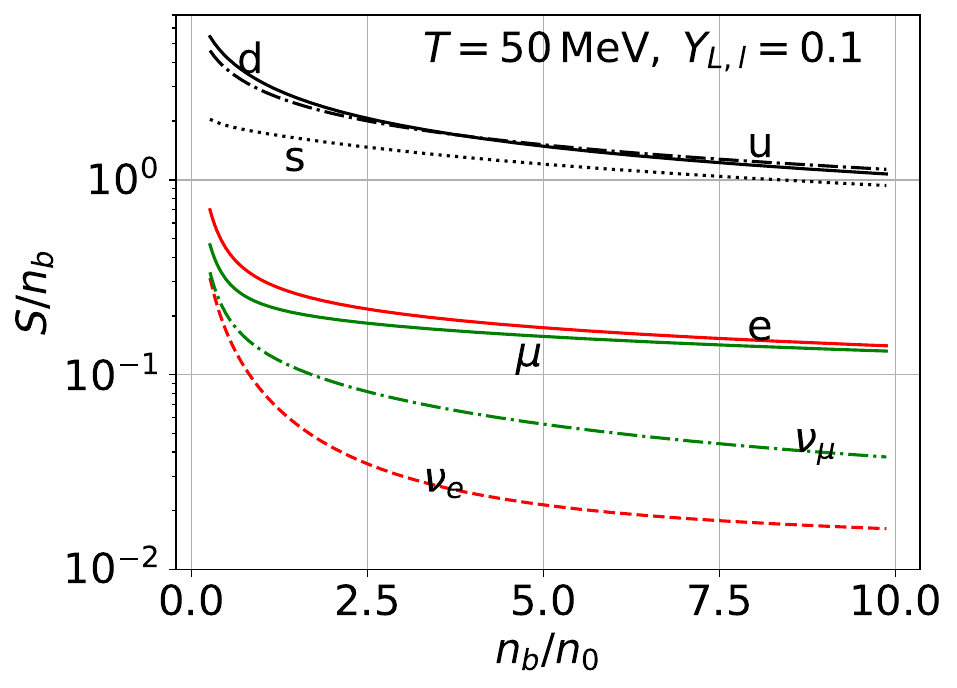}
\caption{In this diagram we show the $Y_i$ and $S/n_b$ as a function of $n_b/n_0$. In the upper panel, we choose two values of $T$; a lower limit of $T=5\,{\rm MeV}$ at which neutrinos are trapped in the remnant matter, and an upper limit $T=50\,{\rm MeV}$, using $Y_{L,e} = Y_{L, \mu} =0.1$ and their corresponding $S/n_b$ distribution for each particle in the bottom panel.}
    \label{pfs}
\end{figure*}
In Fig.~\ref{pfs}, we present the particle distributions, assuming that the NS remnant is at a uniform temperature, using the relation:
\begin{equation}
    Y_i = \dfrac{n_i}{n_u+n_d+n_s},
\end{equation}
where $i$ represents the individual particles present. Comparing the particle distributions in the upper panels, we observe that at low $T$ we have a higher $d$-population relative to $u$ and delayed appearance of the $s$-quark. The $e$'s and $\mu$'s show a marginal increase in the hotter remnant matter. In both cases, the particles do not show any significant changes at higher densities beyond $n_b\sim 2.5n_0$. Also, higher $T$ produces more $\nu_e$ population while $\nu_\mu$ only increases marginally at higher $n_b$ and rather shows a decrease at low $n_b$ with an early appearance of the $s$-quark and $\mu$ in the remnant matter. These effects at higher $T$ cause an increase in ${\rm M_{max}}$ and R as can be seen in Table~\ref{T1}. The isospin asymmetry (or $u-d$ quark asymmetry),
\begin{equation}\label{S}
    \delta = 3 \dfrac{n_d-n_u}{n_d+n_u},
\end{equation}
with $n_3 = n_d-n_u$, the isospin density and $n_b=(n_d+n_u)/3$ the baryon density for two flavor quark systems. The $\delta$ is higher in the low $T$ remnant matter due to excess of $d$-quark relative to $u$ at densities below $n_b \sim 2.7n_0$. In the lower panel, we show the $S/n_b$ contribution of each particle to the overall $S/n_b$. We observe that temperature increase significantly increases the entropy of the particles, as expected, from low $n_b/n_0$ and steadily falls towards the center of the remnant matter. The $d$-quark shows a higher $S/n_b$ contribution while $\nu_e$ shows the least contribution in both cases. %This follows the spread of entropy from the disk towards the remnant core.

\begin{figure}[ht!]
  %\centering
  \includegraphics[scale=0.5]{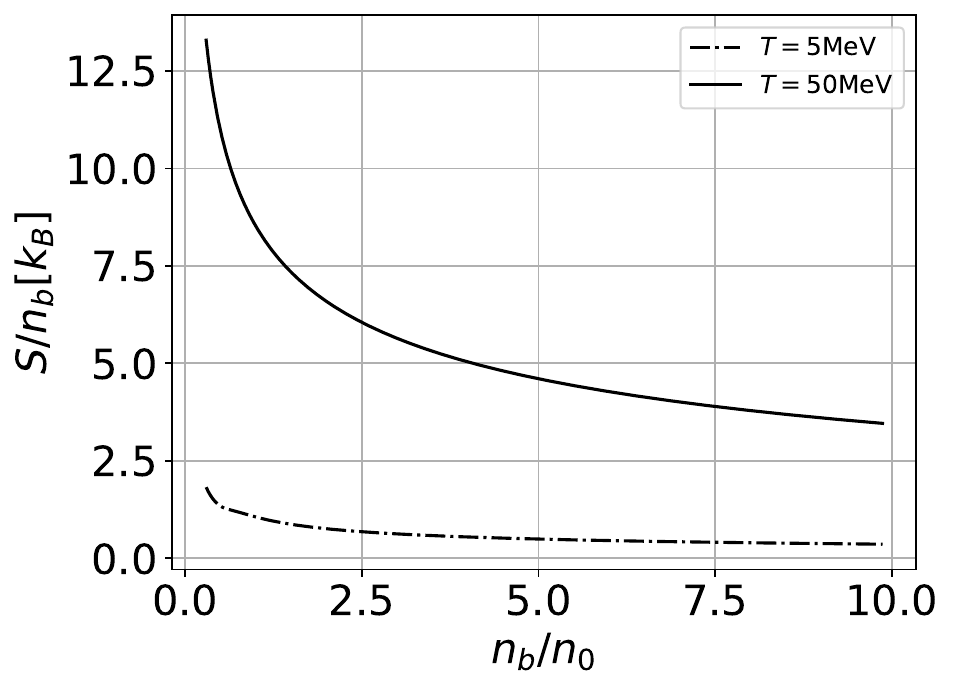}
\caption{Net entropy of the particles per baryon at different fixed temperatures as a function of baryon density. {Solid line for $T=50\,$MeV and dot-dashed line for $T=5\,$MeV.}% TF its is missing the MeV in the figure label!}
}
    \label{ent}
\end{figure}
In Fig.~\ref{ent}, we show the net $S/n_b$ as a function of $n_b/n_0$. We observe a higher $S/n_b$ in hotter QM, as expected. The $S/n_b$ is higher towards the surface (low $n_b$) of the remnant matter, where the disk is formed and decreases toward the center (high $n_b$). The binary BNS merger is associated with net angular momentum and is gravitationally bound; however, thermodynamics in the local rest frame of the system remains significant. In the absence of physical viscosity, the matter flow is isentropic; thus, \( S/n_b \) can serve as a valuable tool for analysis. It has been found in \cite{Perego:2014fma} that the increase in entropy is associated with neutrino absorption, with an initial value in the disk estimated at \( S/n_b \sim 5 - 10 k_B \), rising to \( S/n_b \sim 15 - 20 k_B \) in the later stages of the remnant evolution. Fig.~\ref{pfs} demonstrates that muon neutrinos contribute more significantly to the overall rise in the system's entropy compared to electron neutrinos. %We can deduce that, within the model framework, the \(\nu_e\) contribute significantly more to the rise in entropy compared to the muon neutrinos (\(\nu_\mu\)). Since the muon neutrinos have higher chemical potentials than the electron neutrinos, this leads to a more negative \(\Omega_0\) and a more positive \(S\) for the same temperature \(T\). This relationship can be seen in Eqs.~(\ref{4a}) and (\ref{om}) when considered together.

\begin{figure*}[ht!]
  %\centering
  \includegraphics[scale=0.5]{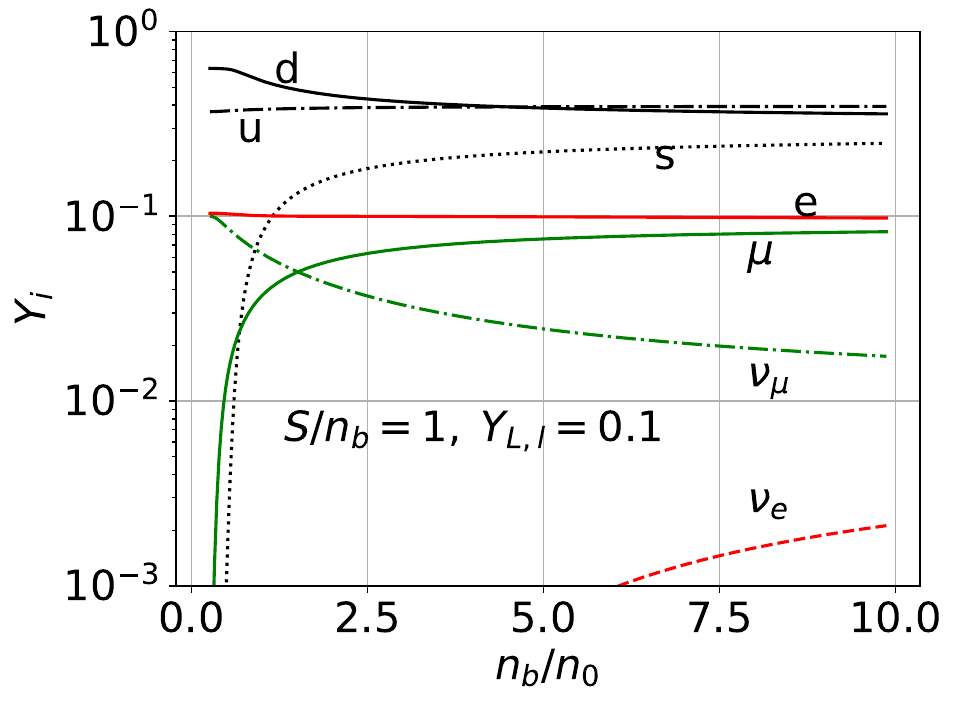}
  \quad
   \includegraphics[scale=0.5]{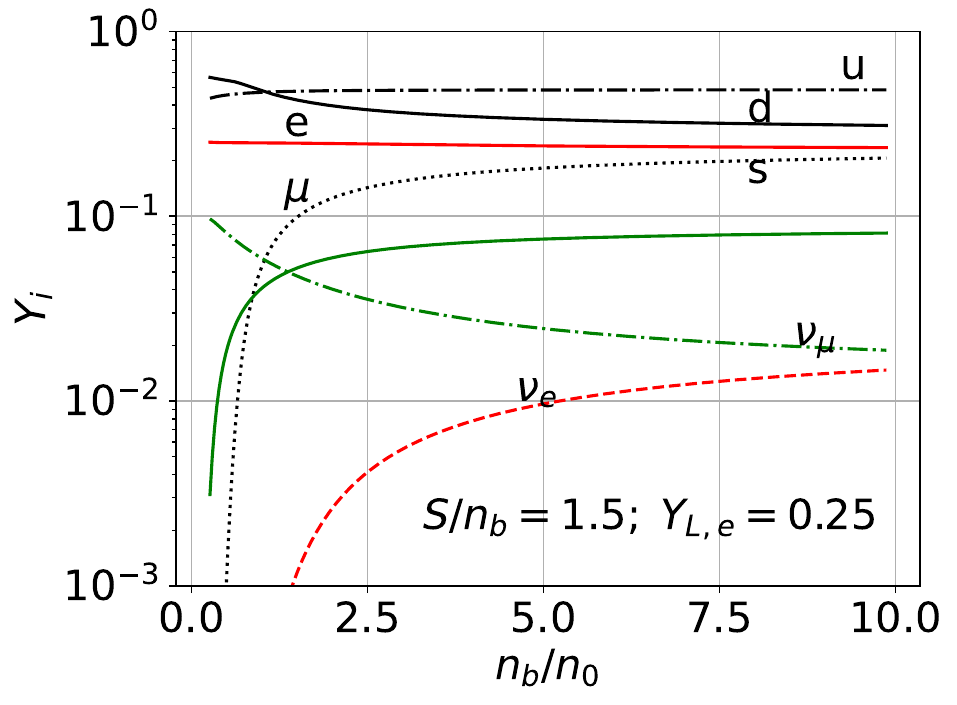}
  \quad
  \includegraphics[scale=0.5]{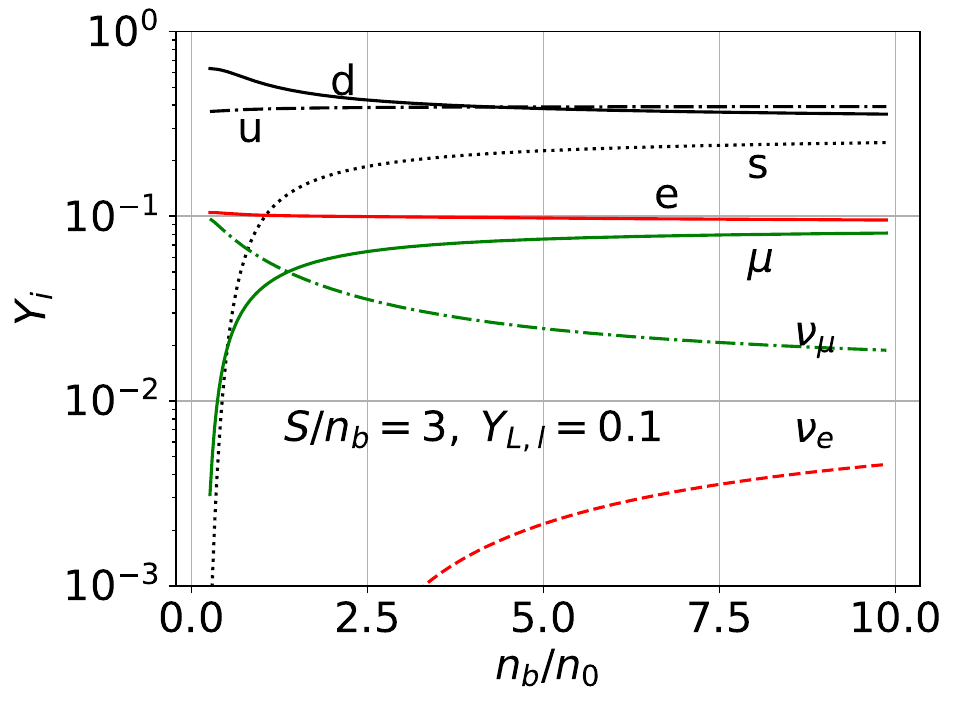}
  \quad
 \includegraphics[scale=0.5]{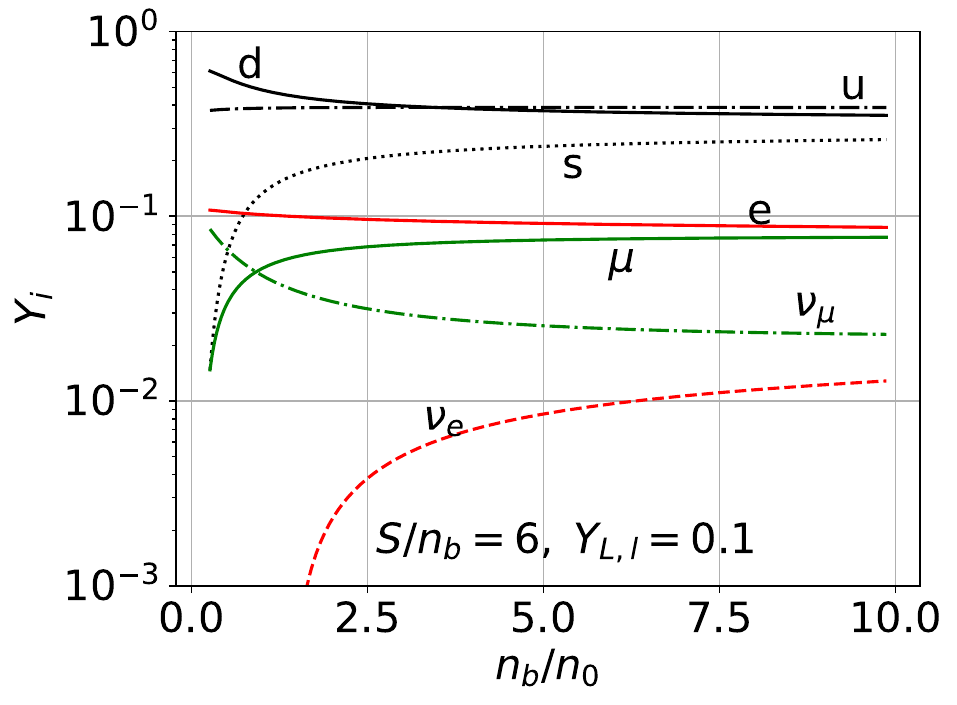}
\caption{We show the $Y_i$ for fixed entropy remnants. The first panel (top left) shows the initial condition at merger $t\sim 0$ ($t$ is time), and the second panel (top right) shows the $Y_i$ in the remnant disk assuming there was prompt collapse of the remnants to a black hole. The bottom panels show two stages after the merger $t>0$. }
    \label{pfs1}
\end{figure*}

In Fig.~\ref{pfs1}, we show four snapshots of the evolution stages of the postmerger NS remnant. The first stage (top left) is the $Y_i$ in the core of the remnant at merger, $t\sim 0$. At this stage, the innermost part of the merging NS cores $n_b>n_0$ does not experience significant shocks, so the $S/n_b$ in the central part of the remnant remains low, it is estimated to be around $1\lesssim S/n_b[k_B ]\lesssim 2$. This stage is characterized by a small temperature profile in the core, mainly generated by matter compression of degenerate nuclear matter above $n_0$. At this stage, we use the initial conditions $S/n_b =1$ and $Y_{L,l} = Y_{L,e} = Y_{L, \mu} = 0.1$ for the analysis, because at $t\sim 0$, the lepton fractions in the cores of the two NSs is frozen to the initial neutrino-less ($\nu$-less) cold weak equilibrium value, which is estimated to be in the range $0.0\lesssim Y_{L,l} \lesssim 0.2$ (this estimate was initially made for only electrons but we generalize it to include muons) towards the higher baryon density regions. The second panel (top right) is based on the assumption of the prompt formation of a black hole. In this case, the thermodynamic conditions change drastically. The formation of an apparent horizon removes all the high-density part of the remnants, leaving a cold disk with an estimated temperature of about $T\lesssim 10\, {\rm MeV}$, with a significant reprocessed lepton fraction of about $Y_{L,l}\,\sim\, 0.25$. At this stage, the disk entropy is slightly higher than when a massive NS is present at the center; hence, we used $S/n_b = 1.5k_B$ as our working assumption. 
Comparing the $Y_i$ in the two top panels, we can see that the right panel has less isospin asymmetry due to an excess of $u$-quark relative to $d$, beginning from $n_b\sim 0.8n_0$ towards higher $n_b$. Also, there is {more} $e$, $\mu$, $\nu_\mu$ and $\nu_e$ content due to higher $Y_{L,l}$. 

At some dynamical timescales, a few milliseconds after the merger, if the remnant does not form a black hole promptly, the inspiral arms' activities at the disk's central part increase the entropy of the remnant matter, creating a strong correlation between the matter density and entropy. The highest temperature of the stream is expected at this stage; however, it may rapidly decrease because of fluid expansion and neutrino diffusion. The lepton fraction is reset to the initial condition due to excess lepton antineutrino emissions and absorption. The entropy at this stage is also estimated to be between $2\leq S/n_b[k_B]\lesssim 10$. In the bottom panels, we set lower and upper limits for $S/n_b$ to study the particle distribution as the $S/n_b$ increases through the evolution of the merger remnant. {Comparing the first panel (bottom left) and the second panel (bottom right), we observe that the NS remnant absorbs more neutrinos as the $S/n_b$ increases, particularly the $\nu_e$ is significantly enhanced while $\nu_\mu$ only increases marginally.} The isospin asymmetry also decreases as $S/n_b$ increases. Comparing the $Y_i$ for the initial condition at merger (top right panel) and the two panels below following the evolution of the NS remnant, assuming it did not form a black hole promptly, the $\nu_e$ significantly increases, and the isospin asymmetry decreases along the evolution lines. It is worth pointing out that, through the evolution stages, the $\nu_e$ shows significant variation at different densities as the NS remnant evolves compared to the other particles. Additionally, higher $S/n_b$ facilitates the emergence of $\mu$ and $s$-quarks, but their effects can be seen at relatively low $n_b$ and do not show any significant variation along the evolution stages at $n_b \approx 1.6n_0$ and above. 

\begin{figure}[ht!]
  %\centering
  \includegraphics[scale=0.5]{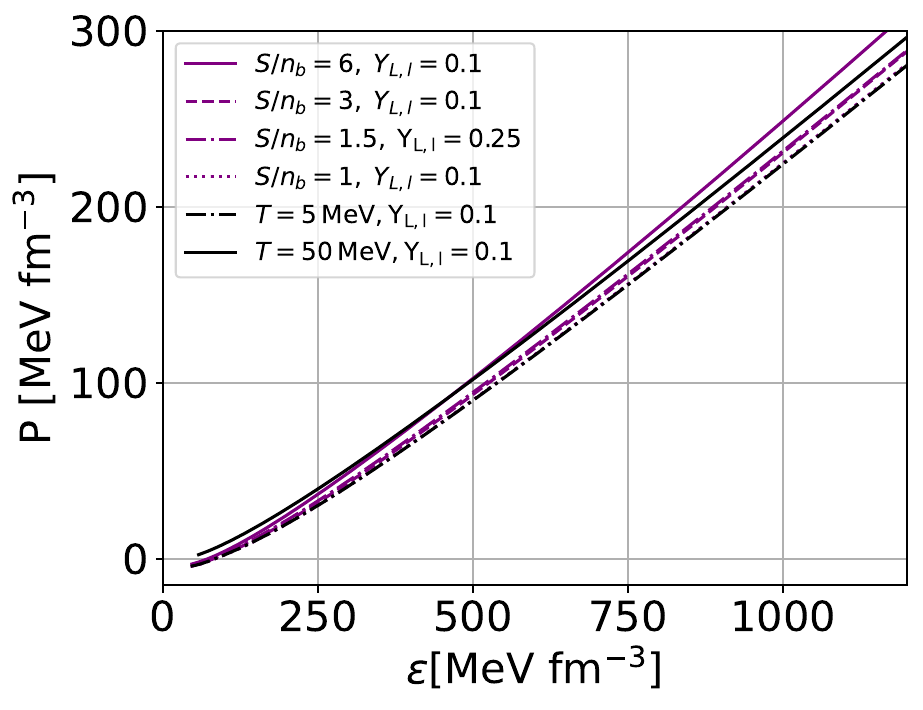}
  % \quad
  %  \includegraphics[scale=0.5]{EoS1.pdf}
\caption{The EoSs for hot quark matter.}
    \label{EoS}
\end{figure}

Fig.~\ref{EoS} shows the EoS of both fixed-entropy NS remnants and uniform-temperature ones. We observe that, in the case of $T=50 {\rm MeV}$, the EoS is stiffer at low $\varepsilon$ and begins to soften at the higher $\varepsilon$ compared to the case of $S/n_b = 6k_B$. This effect can be associated with an increase in the strength of the term proportional to $n_b^{1/3}$ in the equivalent quark mass as $T$ increases. {Comparing the maximum masses and radii data in Table~\ref{T1}, we observe that the fixed-entropy NS remnant is more compact (compactness is determined through $M/R$) than the uniform-temperature one.} Also, comparing their neutrino content, Figs.~\ref{pfs} (top panel, right) and \ref{pfs1} (bottom panel, right), we can see that the $S/n_b = 6k_B$ remnant has a visibly high $\nu_e$ content, particularly toward the center of the remnant. 

Interestingly, as we see in Fig.~\ref{EoS}, the initial conditions for uniformly heated remnant matter at $T = 5{\rm MeV}$ coincide with the initial condition set for fixed entropy remnant matter, although it has a relatively higher core temperature of $T_c\sim 9{\rm MeV}$. Their EoSs overlap each other without distinguishable differences. Comparing their particle fractions, Figs.~\ref{pfs} (top panel, left) and \ref{pfs1} (top panel, left), they have almost the same $\nu_e$ content and early appearance of the $s$-quark, and $\mu$, their $\delta$ are also comparably the same. %Generally, hotter remnant matter has stiffer EoS than its relatively cold counterparts, similarly, 
The remnant matter with higher \( S/n_b \) values has stiffer EoSs compared to those with relatively lower \( S/n_b \), as higher \( S/n_b \) values indicate higher temperatures. In the case of $S/n_b=1.5k_B$, an increase in $Y_{L,l}$ also stiffens the EoS. We can deduce from the particle distribution in the NS remnant matter that a delayed appearance of $\mu,\;\nu_e$ and $s$-quark rather softens the EoS and leads to lower maximum masses and radii. Likewise, a higher $\nu_e$ content in the NS remnant matter leads to a stiffer EoS. Moreover, stiffer EoS are associated with low central energy and baryon densities as shown in Table~\ref{T1}.

\begin{figure}[ht!]
  %\centering
  \includegraphics[scale=0.5]{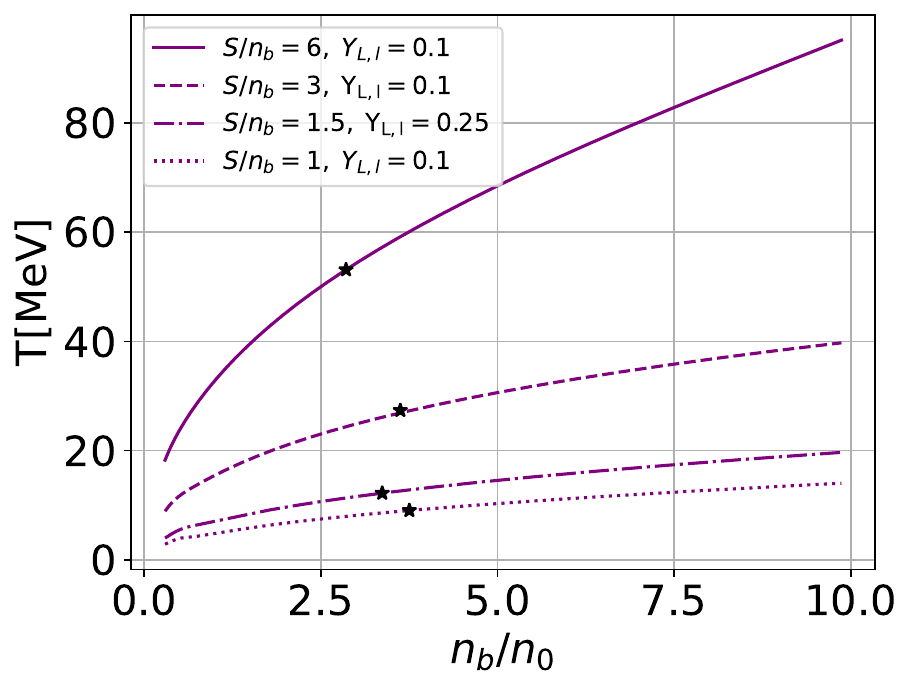}
\caption{The temperature profiles for the fixed entropy remnant matter as a function of the $n_b/n_0$. The black stars in the curves represent the positions of the core temperatures. 
}
    \label{tp}
\end{figure}
In Fig.~\ref{tp}, we show the temperature variations in the NS remnant matter as a function of $n_b/n_0$. At the contact interface of the two NSs, matter from their crusts and outer cores slips out and mixes, causing instability. The cores originally behind these contact interfaces begin to fuse over time, at this point, a series of compressions and expansions occur while the remnants bounce severely. Hence, temperature and density increase immediately after the merger due to these compressions, oscillations, and bounce dynamics \cite{Perego:2019adq, Radice:2020ddv}. Besides, the high sound velocity of the nuclear matter ($c_s  \gtrsim  0.2c$) and nuclear supra-densities ($n_b \gtrsim n_0$) prevent the generation of hydrodynamics shocks in the cores of the two coalescing NSs, thus, the temperature remains low (estimated to be $T\gtrsim 10 {\rm MeV}$) in the core at merger. 

In Fig.~\ref{tp}, we observe that the temperature profile is at its least at the initial merging phase when $S/n_b=1$ ($T_c\sim 9{\rm MeV}$) compared to the other stages when $S/n_b$ was higher (temperature profile for $t\sim 5$ms postmerger has been studied in \cite{Fields:2023bhs} as a function of $n_b$). When the remnant promptly forms a black hole, $S/n_b$ increases, and the central energy and baryon densities decrease, at this point, as can be seen in Fig~\ref{pfs1} for $S/n_b = 1.5k_B$, which corresponds to a core temperature ($T_c\sim 12{\rm MeV}$) and lower central energy and baryon densities relative to the first stage. The evolution process ends when the remnant promptly forms a black hole. However, when the remnant continues evolving as an NS, the cores of the coalescing NSs continue to fuse, causing compression and shear dissipation, increasing the entropy and, as a result, the temperature. We choose a lower limit of $S/n_b =3k_B$ and an upper limit of $S/n_b = 6k_B$ to obtain a fair idea of the temperature variations at this stage of the remnant evolution, we found $T_c\sim 27{\rm MeV}$ and $T_c\sim 53{\rm MeV}$ respectively (higher temperatures in the range of $T \sim 70 - 110 {\rm MeV}$ were determined in \cite{Radice:2020ddv} and $T\sim 158 {\rm MeV}$ reported in \cite{Loffredo:2022prq}). Comparing Figs.~\ref{tp} and \ref{pfs1} we can deduce that higher $S/n_b$ are associated with higher $\nu_e$ production and an early appearance of $\mu$ and $s$-quark in the remnant matter.    

\begin{figure}[ht!]
  \includegraphics[scale=0.5]{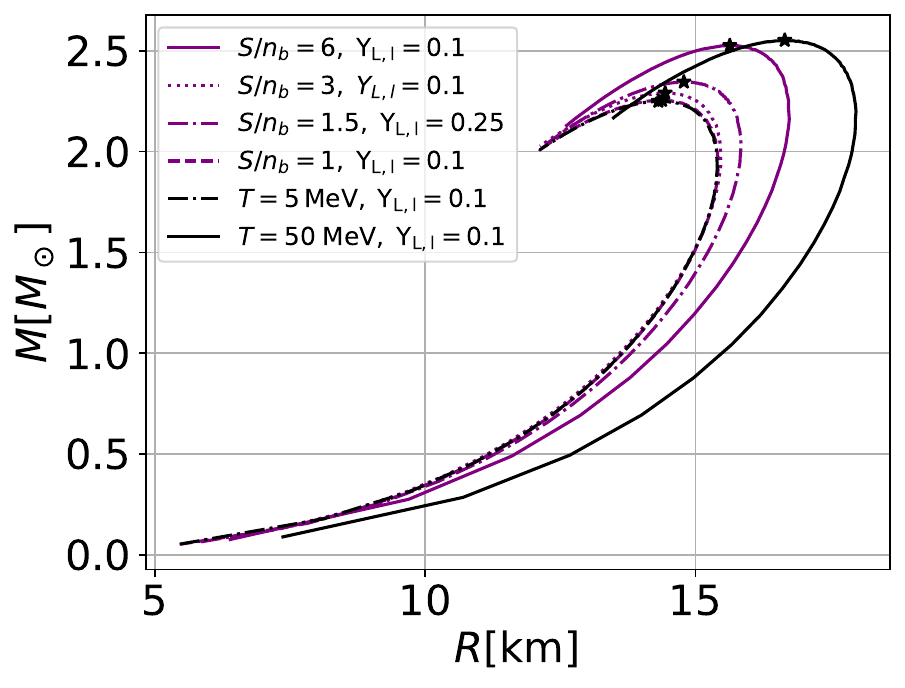}
\caption{The mass-radius diagram of hot star remnant matter. The black dots represent the precise point of the maximum mass and radii of the remnant star. %{\color{red} The red dots have to be increased.} 
}
    \label{mr}
\end{figure}
In Fig.~\ref{mr}, we determine the structure of the NS remnant by calculating the mass-radius (M-R) diagram using the Tolman–Oppenheimer–Volkoff equation (TOV) \cite{PhysRev.55.374} given by the expressions
\begin{align}
    \frac{dP(r)}{dr}&=-[\varepsilon(r) + P(r)]\frac{M(r)+4\pi r^3 P(r)}{r^2-2M(r)r}, \label{eq1}\\
    \frac{dM(r)}{dr}&=4\pi r^2 \varepsilon(r)\label{eq2},
\end{align}
with $r$ the radial coordinate, $M(r)$ the gravitational mass, $P(r)$ the pressure, and $\varepsilon(r)$ the energy density, here, we use the natural units ($G=\hbar =c=1$). {Practically, the NSs whose masses and radii were measured by the NICER observatory data are slowly rotating with estimated spin frequencies of 200\,Hz, however, the computations of rapidly rotating NS models are time-consuming and computationally costly.} Meanwhile, the mass and radii values do not change significantly when slowly rotating rates are assumed and the results are compared to the non-rotating ones. For example, the EoS inferred from the NICER data uses hydrostatic spherically symmetric TOV equations to compare the EoS parameters and the mass-radius values, and they show good agreement~\cite{Miller:2019cac, Raaijmakers:2019qny, Miller:2021qha, Riley:2021pdl}. In this work, we proceed with the assumption that the remnant core did not collapse promptly to a black hole (i.e. a long-lived or infinitely-stable BNS merger remnant core is assumed \cite{Sarin:2020gxb, Bernuzzi:2020tgt}), {so it is characterized by a slow-rotating ``TOV-equivalent" core NS and a rotationally supported disk at large radii in consistency with BNS merger studies}~\cite{Kastaun:2016yaf, Ciolfi:2017uak, Kastaun:2021zyo, Ciolfi:2019fie}. {Thus, the boundary condition is $P(R) = P_{\rm surf}$, where $R$ is the radius of the remnant and $P_{\rm surf}$ is the surface pressure. In the case of cold stars $P_{\rm surf}=0$, on the contrary, $P_{\rm surf}$ is arbitrary. It has been established that the choice of $P_{\rm surf}$ significantly impacts the formation of the mantle at the early stages of stellar evolution. However, significantly low values of $P_{\rm surf}$, which might result from thermally generated pressure, have a negligible effect on the development of the internal structures of the star over a long period \cite{Gulminelli:2015csa, RevModPhys.74.1015}.} It is worth noting that in the two-zone consideration of the binary NS merger (remnant core plus accretion disk), as we discuss here, the total mass of the remnant is concentrated in a slowly rotating dense core, while a small fraction of the rotationally supported mass (the disk) carries the total angular momentum of the system \cite{Margalit:2022rde}. 

The initial condition of the binary NS merger with event GW170817 is two equal-mass NSs with a combined mass estimated to be $2.74^{+0.04}_{-0.01}M_\odot$ and individual masses between 1.17 and 1.6M$_\odot$ \cite{LIGOScientific:2017vwq, Tauris:2017omb, Riley:2019yda}. We neglect the effect of spin in the analysis based on the assumption that most systems are practically non-rotating at merger \cite{1992ApJ...400..175B} together with the arguments about the slow rotation raised above based on a two-zone system. Generally, NSs have a known mass threshold of about 2M$_\odot$ \cite{Fonseca:2021wxt, NANOGrav:2019jur} whereas stellar black hole masses observed in binaries are greater than the GW170817 components \cite{ozel2010black, kreidberg2012mass}. We compare our results with the measured mass and radii of the millisecond pulsar PSR J0740+6620, as determined through observations and analyses by the NICER X-ray observatory. The estimated mass was \( 2.07^{+0.07}_{-0.07} \, {\rm M}_\odot \) and \( 2.08^{+0.09}_{-0.09} \, {\rm M}_\odot \) at a 68\% confidence level (CL), while the radius was measured as \( 12.39^{+1.30}_{-0.98} \, {\rm km} \) \cite{Riley:2021pdl} and \( 13.71^{+2.61}_{-1.50} \, {\rm km} \) \cite{Miller:2021qha}, as determined by different groups. It should be mentioned that the mass of this pulsar was also independently determined to be 2.08$^{+0.07}_{-0.07}{\rm M}_\odot$ through the Shapiro delay effect~\cite{NANOGrav:2019jur}. The mass and radii of PSR J0030+0451 determined through fitting to the NICER data led to two predictions (at 68\% CL); $M = 1.34^{+0.15}_{-0.16}{\rm M}_\odot$, $R =12.71^{+1.14}_{-1.19}{\rm km}$ \cite{Riley:2019yda} and $M = 1.44^{+0.15}_{-0.14}{\rm M}_\odot$, $R = 13.7^{+2.6}_{-1.5}$ km \cite{Miller:2021qha}. Furthermore, NSs with masses $M \sim 2.5 \rm M_\odot$ \cite{LIGOScientific:2020zkf,abbott2020gw190425}  and $M>2.27\rm M_\odot$ \cite{Linares:2018ppq,romani2022psr} have been measured, further imposing stringent constraints on the EoS of NSs and consequently QSs.

It is worth mentioning that our model parameters derived from \cite{daSilva:2023okq}, predicts $R_{1.4} = 14.77$ km (for $S/n_b = 1$ case), which exceeds the commonly quoted inspiral-phase constraint $R_{1.4} \lesssim 13.5$ km derived from LIGO/Virgo analyses \cite{LIGOScientific:2018cki} and Bayesian studies \cite{Annala:2017llu, Dietrich:2020efo}. However, this apparent tension warrants careful context and clarification on three key points: First, the quoted constraints apply to cold, neutrino-transparent NSs in the inspiral phase, as discussed in \cite{Annala:2017llu, Dietrich:2020efo}. In contrast, our study models hot post-merger remnants, characterized by finite temperature, high $S/n_b \geq 1$, and significant populations of trapped electron and muon neutrinos. These conditions contribute substantial thermal pressure (e.g., $P_{\mathrm{th}} \approx 1.5P_{\mathrm{cold}}$ at $S/n_b=1$), leading to a natural expansion of the stellar radius, by up to 10–15\%, as reported in previous studies \cite{Radice:2018pdn}. Second, current observational determinations of the canonical radius $R_{1.4}$ remain subject to significant uncertainties. For example, NICER analyses report $R_{1.4} = 13.7^{+2.6}_{-1.5}$ km \cite{Miller:2021qha} and $12.71^{+1.14}_{-1.19}$ km \cite{Riley:2019yda}, indicating some tension among results even for cold stars. Third, our slightly larger predicted radii arise from the intrinsic stiffening of the EoS of the QM required to support the observed existence of $2 M_\odot$ NS mass, a feature shared by other viable strange or hybrid matter models. Therefore, the $R_{1.4}$ in our model, although somewhat larger, remains consistent with the broader range of uncertainties and is appropriate for the hot and lepton-rich post-merger environments explored in this work.

From our results, the M-R diagram of the initial conditions for fixed $T$ ($T=5{\rm MeV}$) and fixed $S/n_b=1k_B$ overlap, similar to their M$_{\rm max}$'s in Table~\ref{T1}, and are also marked with black stars on the curves. On the other hand, the M-R diagram of $T=50{\rm MeV}$ remnant matter has a larger size with relatively the same M$_{\rm max}$ as $S/n_b=6k_B$ remnant, even though $S/n_b=6k_B$ has a higher core temperature than 50 MeV. Hence, fixed $S/n_b$ NS remnants appear more compact than fixed $T$ NS remnants. Additionally, we observed that the mass and radius of the remnant increase with increasing  
Analysing Figs.~\ref{pfs} and \ref{pfs1} together with the M-R diagram, we can deduce that a lower neutrino content (particularly $\nu_e$) and early appearance of the $\mu$ and $s$-quark in the remnant matter reduces the remnant pressure, thereby softening the EoS and consequently, reducing the maximum mass.  valent quark mass when $T$ increases towards $T_C$.

\begin{figure}[ht!]
  \includegraphics[scale=0.5]{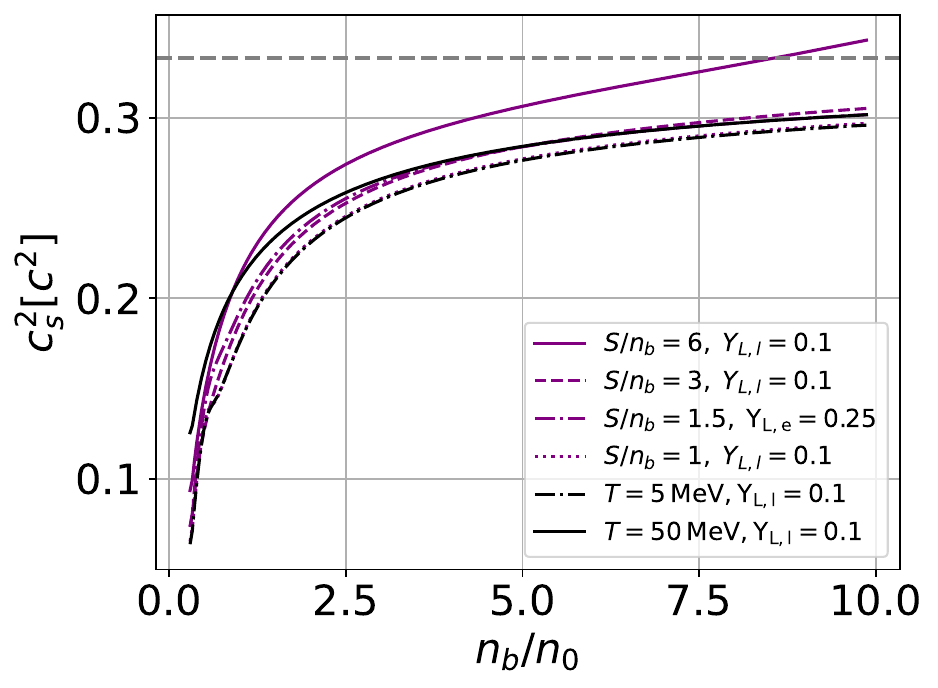}
\caption{The sound velocity through the not star remnant matter as a function of $n_b/n_0$. The horizontal gray line represents the conformal limit, \( c_s^2 = 1/3 \). Above this limit, the matter is considered to violate near-conformality in dense nuclear matter, approaching this limit from below at high $n_b$ means the matter is near-conformal, and it saturates at \( c_s^2 = 1/3 \) in asymptotically free matter reachable at very high densities in pQCD matter.}
    \label{ss}
\end{figure}
In Fig.~\ref{ss}, we show the result of the square of sound velocity, $c_s^2$ as a function of $n_b/n_0$. The $c_s^2$ is determined using the EoS as input from the expression
\begin{equation}
    c_s^2 = \dfrac{\partial P}{\partial \varepsilon},
\end{equation}
where $c_s$ is measured in the unit of constant speed of light ($c$). The $c_s^2$ helps to distinguish between the different phases of strongly interacting matter using conformal symmetry arguments and, more importantly, to measure the stiffness of the EoS. {$c_s^2$ must satisfy the causality constraint $c_s^2\leq 1$ and the thermodynamic stability condition $c_s^2 > 0$ for all kinds of matter compositions \cite{Bedaque:2014sqa}. In exactly conformal matter where the quarks and gluons are in a deconfined state at sufficiently high baryon densities, perturbative QCD (pQCD) predicts that $c_s^2$ saturates at $c_s^2= 1/3$. Again, pQCD predicts that the conformal limit is approached from below in dense nuclear matter \cite{Bedaque:2014sqa, Contrera:2022tqh} before it saturates when the matter is exactly conformal at sufficiently higher densities $n_b\gtrsim 40n_0$ that can only be reached in pQCD \cite{Kurkela:2009gj}. At low densities $n_b\sim 2n_0$, where EoS can be determined with reasonable accuracy using chiral effective field theory with effective degrees of freedom of the pion and nucleon, $c_s^2$ shows a rapid increase with $n_b$ beyond the conformal limit, $c_s^2 = 1/3$ \cite{Leonhardt:2019fua}. Thus, $c_s^2$ is necessary to classify near-conformal matter and matter that violates the conformal constraint (this can be associated with lower degrees of freedom due to the formation of clusters or condensates) in dense nuclear matter at densities that can be reached in NSs or QSs. Consequently, we use the parameter $c_s^2$ to determine the behavior of QM that composes the remnant at each stage of its evolution. The $c_s^2$ was used in determining the quark core in massive NSs (i.e., hybrid NSs) \cite{Annala:2019puf, Issifu:2023ovi} and for determining the behavior of QM that form QSs \cite{daSilva:2023okq}; whether the quarks behave as being confined or free.

While hadronic matter exhibits $c_s^2 > 1/3$ due to nonperturbative effects and chiral symmetry breaking \cite{Bedaque:2014sqa}, dense QM in our DDQM framework shows $c_s^2 \lesssim 1/3$ (except in the case of $S/n_b = 6$) consistent with weakly interacting QM expectations \cite{daSilva:2023okq, Issifu_2025}, though perturbative QCD predicts the conformal limit $c_s^2 = 1/3$ at asymptotically high densities \cite{Braun:2022jme}. We acknowledge that color superconductivity \cite{Contrera:2022tqh, Gartlein:2023vif} and quarkyonic matter scenarios \cite{McLerran:2018hbz} can produce $c_s^2 > 1/3$ through nonperturbative stiffening of the EoS, which may explain massive compact stars \cite{Carlomagno:2023nrc}. Recent extensions of the DDQM model in \cite{Issifu_2025} demonstrate that the formation of bound states through gluon condensation can enhance the stiffness of the EoS while preserving a QM description, thereby producing more massive QSs. In that study, the authors found that heavier QSs tend to violate the conformal limit within the DDQM framework. This suggests that the behavior of $c_s^2$ observed here, although rooted in the microscopic properties of QM, is strongly influenced by the global structure and size of the star. The transition between conformal and non-conformal regimes likely involves complex thermodynamic processes, as discussed in \cite{Ivanytskyi:2022wln, Carlomagno:2024vvr}, particularly those related to degeneracy and chiral symmetry restoration phenomena that warrant further investigation in the context of merger environments. This is evidenced by the deviation of the $c_s^2$ values from the conformal limit at high $S/n_b$, specifically for $S/n_b = 6$.}

The results above show that the $c_s$ is relatively higher in high $S/n_b$ and temperature remnant matter. Almost all the EoSs analyzed for the remnant matter satisfy the near-conformality criteria at higher $n_b$ except $S/n_b = 6k_B$, which shows a slight violation of the conformal limit at high $n_b$ region. In the case of larger sized and less compact remnants (we are comparing $T=50{\rm MeV}$ and $S/n_b= 6k_B$) the $c_s^2$ is relatively higher at low $n_b$ for $T=50$MeV and falls slightly at intermediate to higher $n_b$ without violating the near-conformal criteria, however, $S/n_b = 6k_B$ rises slowly from lower $n_b$ and rises sharply in the intermediate to higher $n_b$ violating the conformal limit at higher $n_b$. As the $S/n_b$ increases, the thermal pressure of the stellar system rises, particularly enhancing the total pressure at moderate $n_b$ where thermal effects are most significant. Although the thermal contribution becomes less dominant at higher densities, the pressure is increasingly governed by degeneracy effects and strong repulsive interactions among particles. In the high-$n_b$ regime, the EoS stiffens due to strong interactions and increased degeneracy, resulting in a rise in the $c_s^2$, with systematically increasing $S/n_b$. Comparing the $c_s^2$ curves to the particle distributions in Figs.~\ref{pfs} and \ref{pfs1}, we can infer that the $\nu_e$ significantly impacts the $c_s^2$ in the remnant matter. The early appearance of $\nu_e$ in the remnant shows higher $c_s^2$ at low $n_b$, and a higher abundance of $\nu_e$ toward the center of the remnant results in higher $c_s^2$ at higher $n_b$.

\begin{figure}[ht!]
  \includegraphics[scale=0.5]{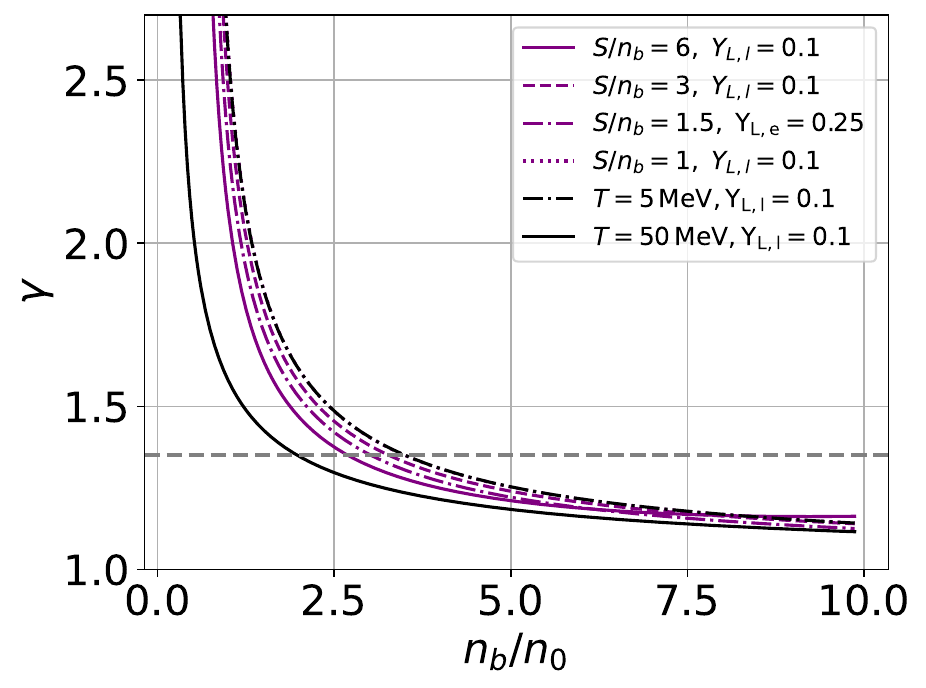}
\caption{The polytropic index as a function of $n_b/n_0$. The horizontal gray line is the conformal limit, $\gamma = 1.35$, which we determine by taking an average of the range of values reported in \cite{Kurkela:2009gj, Gorda:2021znl} for pQCD matter $\gamma = [1, 1.7]$.}
    \label{plt}
\end{figure}
In Fig.~\ref{plt}, we show the polytropic index as a function of $n_b/n_0$ determined through the relation
\begin{equation}
    \gamma = \dfrac{\partial\ln P}{\partial \ln\varepsilon},
\end{equation}
using the EoS as input. The $\gamma$ is also used to classify the inner composition and the phases of matter in the NS remnant matter. The application of $\gamma$ is similar to $c_s$ discussed above. They constitute the two most important parameters for determining the classes of matter that form an NS or exist in the core of the NS (if the existence of a hybrid NS is assumed \cite{Annala:2019puf}) using the EoS. A conformally symmetric matter is scale-independent and does not involve any dimensionful parameter; therefore,  $P$ and $\varepsilon$ are expected to be proportional, given rise to $\gamma =1$. On the contrary, the value of $\gamma$ in dense nuclear matter with hadronic degrees of freedom is estimated to be $\gamma \approx 2.5$ around and above $n_0$ in chiral effective theories with nucleon and pion degrees of freedom, setting a threshold for $\gamma$ in hadronic matter \cite{Kurkela:2009gj, Liu:2023gmq}. This is expected because the ground state of QCD does not have its usual approximate chiral symmetry at low to intermediate baryon densities, $n_b \sim 2n_0$ \cite{Holt:2009ty}. From Fig.~\ref{plt}, the $\gamma$ decreases as $n_b/n_0$ increases and attains its minimum value $\gamma \approx 1.1$. This value is slightly above the conformal limit; however, it is in good agreement with the approximate conformal matter classification range in pQCD \cite{Kurkela:2009gj, Gorda:2021znl} and the results determined in \cite{Annala:2019puf} after examining a wide range of EoSs, which estimates $\gamma \leq 1.75$. Moreover, we observed that heavier remnants approach the $\gamma$ threshold faster from above $\gamma > 1$ compared to less massive remnants.

\section{Final remarks}\label{fr}
We studied QSs under the thermodynamic conditions characteristic of BNS postmerger remnants in the presence of $e$ and $\mu$ and their corresponding neutrinos, assuming that the merger remnant is a compact star composed of QM, using the DDQM model. The study of merger remnants has generally been investigated with the assumption that the binary system is composed of hadronic stars. Therefore, this study aims to contribute to the perspective that the binary system could potentially be composed of strange quark stars \cite{Zhou:2021tgo, Zhou:2024syq}. Generally, the aftermath of the merger has three different outcomes: 1) it may form a stable NS, 2) form a black hole, or 3) form a  supramassive rotationally supported NS, which will later collapse to form a black hole due to angular momentum losses. The possibility of any of these three scenarios occurring depends on the masses and the EoSs of the binaries participating in the merger event. In each scenario, the event will affect the GW signal and its electromagnetic counterpart (see e.g. Refs.~\cite{Sarin:2020gxb, Piro:2017zec, Radice:2020ddv} for more detailed discussions). Assuming a stable NS (in this case a QS) was formed in the framework of this study, a few seconds after the merger, the remnant will go through deleptonization, neutrino diffusion, and thermal radiation to form a stable SQS at $T=0$, several years later. In that case, the $M_{\rm max}=2.18{\rm M}_\odot$ and the radius $R=13.86{\rm km}$ as shown on Table~II of \cite{daSilva:2023okq}. 

This work focuses on introducing temperature into the DDQM model using a lattice QCD–motivated approach, for the first time, while also consistently accounting for the presence of $\nu_\mu$ and $\mu$ in the composition of the merger remnant throughout its evolution process. The main findings of our investigation are:
\begin{itemize}
    \item In the upper panels of Fig.~\ref{pfs} and in Fig.~\ref{pfs1} we observed a significant increase in $\nu_e$ as the merger remnant evolves with increasing $T$ and $S/n_b$. The $\nu_e$'s shift more towards the low $n_b$ region, and their population also increases significantly toward the center of the remnant matter along the evolution lines. The $\nu_\mu$ population drops marginally at low $n_b$ and does not show any significant increase at higher $n_b$ along the evolution lines. The $\delta$ of the remnant matter also decreases with increasing $T$ and $S/n_b$ due to an increase of $u$-quark over $d$-quark at relatively high $n_b$ along the evolution lines. As expected, an increase in $T$ and $S/n_b$ facilitates the early appearance of $s$-quark in the remnant matter; a similar effect was observed with the $\mu$'s. Particle distribution under conditions of merger involving hadronic matter can be found in \cite{Sedrakian:2021qjw, Sedrakian:2022kgj}; their results qualitatively agree with our findings, considering that neutrons are rich in $d$, protons are rich in $u$, and hyperons contain $s$-quarks. 

    \item In the lower panels of Fig.~\ref{pfs}, we found that massless particles like $\nu_e$ and $\nu_\mu$ contribute less to the overall $S/n_b$ of the system. Generally, an increase in $T$ mutually increases the $S/n_b$ of each particle in the system, thereby increasing the net $S/n_b$ as shown in Fig.~\ref{ent}. Additionally, the $S/n_b$ decreases with increasing $n_b$ among the individual particles; a similar effect is observed in the net system of particles in Fig.~\ref{ent}. This reinforces the point that the $S/n_b$ spreads from the disk towards the center of the remnants (higher $S/n_b$ in the time interval of $\sim 40{\rm ms}$ postmerger have been reported in \cite{Perego:2014fma}).

    \item In Fig.~\ref{EoS}, we found that a delayed appearance of the $s$-quark and $\mu$ in the remnant matter coupled with a large $\delta$ soften the EoS resulting into lower M$_{\rm max}$ and $R$. On the other hand, a shift of the $\nu_e$ towards the low $n_b$ region increases the pressure in the remnant matter and stiffens the EoS, leading to higher M$_{\rm max}$ and $R$. The $\nu_\mu$ constituent does not significantly affect the EoS along the evolution lines; however, its marginal decrease at low $n_b$ and marginal increase at higher $n_b$ coincides with the stiffening of the EoS as the remnant evolves. Generally, we observed that the remnants' EoS becomes stiffer as $T$ and $S/n_b$ increase, as a result of the $n_b^{1/3}$ dependence of the equivalent quark mass and the enhancement of this term by driving $T$ to the critical deconfinement temperature.

    \item We observed a connection between temperature rise, neutrino production, and the emergence of different particles in the remnant matter. The matter with the highest temperature profile in Fig.~\ref{tp} corresponds to the matter with the highest neutrino content, particularly $\nu_e$, early appearance of $s$ and $\mu$, and less $\delta$. We find that the temperature increases with $n_b$ and higher values of $S/n_b$, reaching core temperatures ranging between $T_c\sim 9-53{\rm MeV}$ along the evolution lines considered. These values are within the lower limits reported in Refs.~\cite{Loffredo:2022prq, Perego:2019adq, Most:2022wgo} for postmerger remnant matter. Temperature increases in the remnant reduce both the central energy and central baryon densities. Temperature fluctuations in quark matter are generally lower than the corresponding hadronic matter investigated under the same conditions due to higher degeneracy in quark matter.

    \item The EoSs calculated and analyzed produced non-rotating spherically symmetric static compact objects with maximum masses of between $2.25-2.55$M$_\odot$, which is fairly in agreement with the estimated combined masses of the NS binaries that were involved in the GW170817 event \cite{LIGOScientific:2017vwq} and the secondary component in the GW190814 event with maximum mass of $2.59^{+0.08}_{-0.09}$M$_\odot$ \cite{LIGOScientific:2020zkf}. The maximum masses also satisfy the threshold set by PSR~J0740+6620, and the remnant radii determined are within the upper limits of the radius determined for PSR J0740+6620 by \cite{Miller:2021qha} through the analyses of the NICER observatory data (see the result in Fig.~\ref{mr} and Table~\ref{T1}). Since the maximum masses obtained here are higher than the ones obtained for PSR~J0740+6620, even though their radii correspond to the upper limit, we can infer that the remnants analyzed here are more compact. Besides, the results satisfy the maximum NS mass constraints set through more massive PSR~J2215+5135~\cite{Linares:2018ppq} and PSR~J0952-0607~\cite{Romani:2022jhd} observed pulsars.

    \item Finally, we analyzed the composition of the remnant matter through the conformal symmetry arguments using the EoS as input. The results obtained for $c_s$ and $\gamma$ are presented in Figs.~\ref{ss} and \ref{plt} respectively. The graphs show $c_s^2<1/3$ except for $S/n_b = 6k_B$, and $\gamma\gtrsim 1.1$, in conformity with near-conformal matter classification detailed in \cite{Annala:2019puf} through a wide range of EoSs analyzed.
\end{itemize}
The thermally generated neutrinos observed along the evolution lines significantly influence the stiffness of the EoS and consequently the other properties of the remnant evolution calculated through the EoSs. 

\section{Acknowledgements}
A.I. thanks the  financial support from the
 S\~ao Paulo State Research Foundation (FAPESP) Grant No.  2023/09545-1. T.F. is supported by the National Council for Scientific and Technological Development (CNPq) under Grants Nos. 306834/2022-7. T. F. also thanks the financial support from  Improvement of Higher Education Personnel CAPES (Finance Code 001) and FAPESP Thematic Grants (2019/07767-1 and 2023/13749-1). This work is a part of the
project Instituto Nacional de  Ci\^{e}ncia e Tecnologia - F\'{\i}sica
Nuclear e Aplica\c{c}\~{o}es  Proc. No. 464898/2014-5.

\bibliographystyle{ieeetr}
\bibliography{references.bib}

\end{document}